\documentclass[9pt,twocolumn,twoside]{osajnl}
\journal{pr} 
\setboolean{shortarticle}{false}
\input{glyphtounicode}
\pdfglyphtounicode{parenleftbig}{0028}
\pdfglyphtounicode{parenrightbig}{0029}
\pdfglyphtounicode{parenleftBig}{0028}
\pdfglyphtounicode{parenrightBig}{0029}
\pdfglyphtounicode{parenleftbigg}{0028}
\pdfglyphtounicode{parenrightbigg}{0029}
\pdfglyphtounicode{parenleftBigg}{0028}
\pdfglyphtounicode{parenrightBigg}{0029}
\pdfglyphtounicode{bracketleftbig}{005B}
\pdfglyphtounicode{bracketrightbig}{005D}
\pdfglyphtounicode{bracketleftbigg}{005B}
\pdfglyphtounicode{bracketrightbigg}{005D}
\pdfglyphtounicode{bracketleftBigg}{005B}
\pdfglyphtounicode{bracketrightBigg}{005D}
\pdfglyphtounicode{braceleftbig}{007B}
\pdfglyphtounicode{bracerightbig}{007D}
\pdfglyphtounicode{braceleftbigg}{007B}
\pdfglyphtounicode{bracerightbigg}{007D}
\pdfglyphtounicode{braceleftBigg}{007B}
\pdfglyphtounicode{bracerightBigg}{007D}
\usepackage{xcolor}
\usepackage{balance}
\title{Learning Transferable Self-Supervised Priors for Super-Resolution Reconstruction in Structured Illumination Microscopy}

\author[1,2,3]{Tong-Tian Weng}
\author[1,2,3,*]{Ze-Hao Wang}
\author[1,2,3]{Qi Wang}
\author[1,2,3]{Xi-Hua Wang}
\author[1,2,3,4]{Xiang-Dong Chen}
\author[1,2,3,4,*]{Fang-Wen Sun}

\affil[1]{Laboratory of Quantum Information, University of Science and
Technology of China, Hefei 230026, China}
\affil[2]{CAS Center for Excellence in Quantum Information and Quantum Physics, University of Science and Technology of China, Hefei 230026, China}
\affil[3]{Anhui Province Key Laboratory of Quantum Network, University of Science and Technology of China, Hefei 230026, China}
\affil[4]{Hefei National Laboratory, University of Science and Technology of China, Hefei 230088, China}
\affil[*]{Corresponding authors: fwsun@ustc.edu.cn, zehao@ustc.edu.cn}

\begin{abstract}

Structured illumination microscopy (SIM) extends the optical passband, while recovering detail beyond it depends on structural priors that can degrade when imaging conditions or specimens change. We introduce SIMAdapter, a two-stage self-supervised framework without paired high-resolution references. It first pretrains a shared emitter and point-spread function (PSF) network on 23,585 raw SIM stacks from BioSR, BioTISR, and simulations with diverse PSFs and structures, then adapts it to a single unlabeled target stack against a calibrated differentiable image-formation model. Across two held-out synthetic domains, SIMAdapter reaches a mean emitter normalized root-mean-square error of 0.144 after 1200 updates, compared with 0.403 for Sparse-SIM and 0.315 for the most accurate published learned baseline, whereas the same adaptation from random weights or BioSR-only pretraining reaches 0.528 and 0.172. Adaptation also reduces error across the tested 0.10-wave pupil aberrations and simulated changes in numerical aperture, pixel size, and illumination period. On an experimental Argolight calibration target, it preserves flanking single reference lines as single peaks while resolving central double line pairs at nominal 120, 90, and 60 nm spacings. A single pretrained network can thus be reused across SIM measurements, with each reconstruction refined against its own raw data.

\end{abstract}

\makeatletter
\renewcommand{\p@subsection}{\thesection.}
\renewcommand{\p@subsubsection}{\thesection.}
\makeatother

\setboolean{displaycopyright}{true}
\doi{}

\begin{document}
\maketitle

\section{Introduction}

Structured illumination microscopy (SIM) combines super-resolution imaging with conventional fluorophores and rapid acquisition~\cite{wu2018faster,demmerle2017strategic}. By shifting otherwise inaccessible spatial frequencies into the detectable passband, the light pattern allows linear SIM to approximately double the resolution of wide-field microscopy~\cite{gustafsson2000sim,chen2023simreview}. Recovering these frequencies, however, requires an accurate image-formation model: errors in the assumed light pattern, point-spread function (PSF), background, or noise statistics can obscure fine structures and introduce artifacts~\cite{wicker2013phase,smith2021structured,ball2015simcheck,pospivsil2017analysis}. Methods such as PCA-SIM, HiFi-SIM, Open-3DSIM, and accelerated nonlinear SIM reconstruction frameworks therefore improve reconstruction through pattern estimation and optical modeling~\cite{qian2023structured,wen2021high,cao2023open,zhang2025high}.

Beyond the conventional SIM passband, reconstruction depends increasingly on prior knowledge of the specimen, which hand-designed regularizers encode as structural assumptions. Hessian-SIM regularizes low-photon reconstructions with a Hessian penalty~\cite{huang2018fast}, and Sparse-SIM combines sparsity and continuity to recover detail beyond the conventional SIM limit~\cite{zhao2022sparse}.

Learned priors instead encode structural patterns from examples. DL-SIM, ML-SIM, and DFCAN use supervised networks to improve reconstruction under challenging acquisition conditions~\cite{jin2020deep,christensen2021ml,qiao2021evaluation}, rDL-SIM combines a supervised network with the SIM image-formation model to denoise the raw frames before reconstruction~\cite{qiao2023rationalized}, and deep-learning noise correction has likewise been developed for fluorescence microscopy~\cite{qu2025deep}. Public datasets such as BioSR support the training and evaluation of such methods~\cite{qiao2021evaluation}. Supervised training, however, requires paired high-resolution references or simulated targets, which are hard to obtain across microscopes, light patterns, noise levels, and specimens. A trained network may also perform worse when these conditions change~\cite{belthangady2019applications}.

Physics-based self-supervision removes the need for paired references by training a network against the raw measurements themselves through an image-formation model. The network predicts the emitter, and a differentiable forward model maps that prediction back to the measurement domain, where it is compared with the acquired data. This approach has enabled reference-free reconstruction in light-field microscopy~\cite{lu2025physics} and self-supervised volumetric restoration in two-photon microscopy~\cite{li2025self}, and has recently been explored for SIM. SSR-SIM draws on structured-light modulation priors and reconstruction artifact statistics~\cite{liu2025bio}, and in our previous work we introduced a network that predicts both the emitter and the PSF from raw SIM data and trained it on BioSR~\cite{wang2025self}. SSR-SIM and our previous network were each trained under one set of optical conditions. In practice, however, SIM instruments vary across laboratories and imaging sessions in numerical aperture, pixel sampling, emission wavelength, illumination period, optical aberrations, and specimen structure, where collecting a separate dataset and retraining a model for each configuration is rarely practical. Endowing a self-supervised prior with transferability across optical conditions and single-stack adaptability is therefore essential for making learned SIM reconstruction a reusable tool across the microscopy community.

\begin{figure*}[!t]
    \centering
    \includegraphics[width=\textwidth]{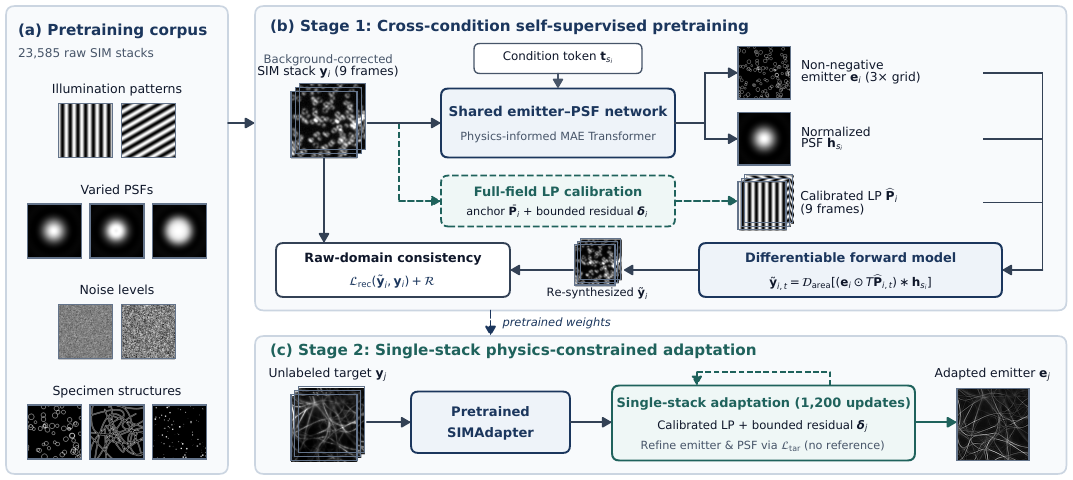}
    \caption{
        \textbf{Learning and adapting a shared SIM reconstruction prior.} (a) \emph{Pretraining corpus} combining \(23{,}585\) raw SIM stacks from BioSR, BioTISR, and simulations spanning varied illumination patterns, PSFs, noise levels, and specimen structures. (b) \emph{Stage 1: cross-condition self-supervised pretraining} jointly trains a shared emitter--PSF network conditioned on discrete tokens \(\mathbf{t}_{s_i}\); a differentiable forward model combines the predicted emitter \(\mathbf{e}_i\), normalized PSF \(\mathbf{h}_{s_i}\), and calibrated light pattern \(\widehat{\mathbf{P}}_i\) (anchor \(\bar{\mathbf{P}}_i\) plus bounded residual \(\boldsymbol{\delta}_i\)) to match the acquired stack \(\mathbf{y}_i\) through raw-domain consistency \(\mathcal{L}_{\mathrm{rec}}\) and physical regularizers \(\mathcal{R}\) without paired high-resolution references. (c) \emph{Stage 2: single-stack physics-constrained adaptation} transfers the pretrained weights and refines the emitter, PSF, and bounded light-pattern residual \(\boldsymbol{\delta}_j\) on a single unlabeled target stack \(\mathbf{y}_j\) for \(1200\) updates under \(\mathcal{L}_{\mathrm{tar}}\).
    }
    \label{fig:pipeline}
\end{figure*}

SIMAdapter addresses this need in two stages (Fig.~\ref{fig:pipeline}). It keeps the physics-informed Transformer architecture and the joint emitter--PSF formulation of our previous network~\cite{wang2025self}; its contribution is to turn a single-condition self-supervised network into a transferable, adaptable reconstruction prior that can be reused directly on new measurements. Pretraining on raw SIM data from BioSR, BioTISR, and simulations with varied PSFs, with illumination phases estimated from each stack and group-consensus carriers used when experimental estimates are unstable, learns one prior shared by many acquisitions. One-shot adaptation then specializes this prior to a single unlabeled target stack through the calibrated forward model, refining the PSF from the target data. We test which parts of this procedure produce the gain: controlled comparisons separate pretraining from adaptation and isolate the PSF, the updated parameters, and the dark-region penalty; tests under unseen aberrations and pixel-space PSF scales probe transfer across optical conditions; and published supervised and self-supervised networks provide references on the same data. On two held-out synthetic domains, the adapted reconstructions are more accurate than the pretrained network's direct predictions, the classical references, and these learned methods. On three experimental FlexSIM measurements~\cite{soubies2024surpassing}, adaptation removes artifacts of the direct predictions, and on an Argolight calibration target it distinguishes flanking single reference lines from central double line pairs across the tested nominal \(120\), \(90\), and \(60~\mathrm{nm}\) spacings, with or without the dark-region penalty.


\section{Methods}

\subsection{Reconstruction network and forward model}
\label{sec:recon}\label{sec:arch}\label{sec:psf-cond}

SIMAdapter learns an emitter--PSF model by matching forward-model predictions to the acquired SIM frames. The emitter network uses the physics-informed Transformer architecture of our previous work~\cite{wang2025self}, pretrained as a physics-informed masked autoencoder~\cite{he2022masked,feichtenhofer2022masked,wang2024learning}, and predicts a non-negative emitter intensity distribution on a threefold-upsampled grid. Each raw stack carries a source-level optical-condition index \(s_i\), and a PSF branch maps the token of that condition to a non-negative, unit-sum kernel \(\mathbf{h}_{s_i}\) through a spatial softmax. All conditions share the model parameters, and each condition is identified by its own distinct token. Appendices~\ref{app:psfcond} and~\ref{app:protocol} describe the conditioning and the architecture.

For a raw stack \(\mathbf{x}_i=\{\mathbf{x}_{i,t}\}_{t=1}^{T}\), with \(T=9\), background correction and normalization give \(\mathbf{y}_i=\mathcal{N}[\mathcal{B}(\mathbf{x}_i)]\). Full-field calibration with the FlexSIM procedure~\cite{soubies2024surpassing} (using acquisition-group or training-cell carrier consensus when experimental carrier estimates are unstable, Appendix~\ref{app:calibration}) provides a light-pattern anchor, and a bounded residual \(\boldsymbol{\delta}_i\) adjusts its gains, phases, and wave vectors. The adjusted light pattern is evaluated at the sample centers of the threefold grid in detector coordinates, the convention used to render the calibrated anchor, with the frame gains constrained to sum to \(T\). The forward model uses this calibrated light pattern and the background-corrected measurements in place of the light-pattern and background heads of the original architecture, which are bypassed.

The predicted emitter \(\mathbf{e}_i\), PSF \(\mathbf{h}_{s_i}\), and residual-adjusted light pattern \(\widehat{\mathbf{P}}_i\) then produce frame \(t\) through
\begin{equation}
    \tilde{\mathbf{y}}_{i,t}
    =
    \mathcal{D}_{\mathrm{area}}
    \left[
        \left(\mathbf{e}_i\odot T\widehat{\mathbf{P}}_{i,t}\right)
        \ast\mathbf{h}_{s_i}
    \right],
    \label{eq:sim_forward_frame}
\end{equation}
where \(\tilde{\mathbf{y}}_{i,t}\) is the predicted frame, \(\odot\) denotes modulation by the light pattern, \(\ast\) denotes convolution, and \(\mathcal{D}_{\mathrm{area}}\) integrates over detector pixels to return the prediction to the measurement grid. The factor \(T\) converts the solver's light-pattern convention to the forward-model convention (Appendix~\ref{app:normalization}). The reconstruction loss measures the agreement between the predicted stack \(\tilde{\mathbf{y}}_i\) and the measured stack \(\mathbf{y}_i\); no paired emitter or PSF reference is required.

\subsection{Learning across optical conditions}
\label{sec:corpus}\label{sec:pretrain}\label{sec:objective}

Pretraining uses \(23{,}585\) raw SIM stacks from BioSR, BioTISR, and simulations spanning different PSFs and specimen structures. All FlexSIM target measurements are excluded. Appendix~\ref{app:corpus} gives the corpus composition, data splits, and sampling procedure.

Starting from the checkpoint of our previous network, trained on BioSR alone~\cite{wang2025self}, we train the model on this mixed corpus. The objective for each stack combines raw-measurement consistency with regularization of the optical model:
\begin{equation}
\begin{aligned}
    \mathcal{L}^{(i)}
    ={}&\mathcal{L}_{\mathrm{rec}}^{(i)}
    +\lambda_h\mathcal{R}_{\mathrm{grad}}(\mathbf{h}_{s_i})
    +\lambda_P\mathcal{R}_{\mathrm{grad}}(\widehat{\mathbf{P}}_i)\\
    &+\lambda_c\mathcal{R}_{\mathrm{center}}(\mathbf{h}_{s_i})
    +\lambda_{\delta}\mathcal{R}_{\delta}(\boldsymbol{\delta}_i).
\end{aligned}
\label{eq:training_loss}
\end{equation}
Here \(\mathcal{L}_{\mathrm{rec}}^{(i)}\) is the reconstruction term, which combines a masked raw-domain \(\ell_1\) error between \(\tilde{\mathbf{y}}_i\) and \(\mathbf{y}_i\) with a loss based on 3D multiscale structural similarity (MS-SSIM). The term \(\mathcal{R}_{\mathrm{grad}}\) is a mean squared finite-difference penalty over the represented dimensions of the mean-normalized PSF \(\mathbf{h}_{s_i}\) and light pattern \(\widehat{\mathbf{P}}_i\); \(\mathcal{R}_{\mathrm{center}}\) penalizes the displacement of the PSF centroid from the center of its support; and \(\mathcal{R}_{\delta}\) penalizes the magnitude of the bounded residual \(\boldsymbol{\delta}_i\), that is, departures from the calibrated light-pattern anchor. The weights are \(\lambda_h=\lambda_P=\lambda_\delta=10^{-3}\) and \(\lambda_c=10^{-1}\); Appendix~\ref{app:objective-full} gives the full expressions. Pretraining jointly optimizes the emitter network, PSF branch, and per-stack light-pattern residuals while keeping the condition tokens fixed. Appendix~\ref{app:protocol} describes the architecture, augmentation, and optimization procedure.

\subsection{Adaptation to a single measurement}\label{sec:adapt}

Starting from the shared pretrained checkpoint, each run adapts the model to one complete, unlabeled nine-frame stack and reconstructs its underlying emitter. We apply background correction followed by full-stack percentile normalization (\(0.01\mathrm{st}\) to \(99\mathrm{th}\) percentiles, Appendix~\ref{app:adapt-obj}) before adaptation. Full-field calibration supplies the target light-pattern anchor. Adaptation updates the emitter network, PSF branch, and bounded light-pattern residual; the condition tokens remain fixed.

Adaptation uses the pretraining forward model and regularizers with masked reconstruction disabled. It adds an emitter-intensity penalty weighted toward darker measurement regions, with weight \(\lambda_d=3\times10^{-3}\), and raises the PSF-gradient weight to \(\lambda_h=2\times10^{-3}\). Appendix~\ref{app:adapt-obj} defines the adaptation loss terms. No high-resolution reference is used for parameter updates or checkpoint selection. All primary adaptation comparisons use a fixed budget of \(1200\) updates, and the optimizer settings are specified in Appendix~\ref{app:adapt-settings}.

\subsection{Evaluation}
\label{sec:baselines}\label{sec:metrics}\label{sec:decorr}

We compare adapted reconstructions with a zero-shot reference that applies the same pretrained checkpoint in a single forward pass without a target update. Both outputs use the same threefold reconstruction grid. Our previous network, trained on BioSR alone, supplies the initialization. Section~\ref{sec:adaptation_results} evaluates it, a randomly initialized network, and ablations of the adaptation procedure under the same settings.

The classical references are wide-field averaging, conventional SIM, FlexSIM, and Sparse-SIM. Conventional SIM receives the simulated light-pattern geometry and transfer function on synthetic data, and uses blind FlexSIM reconstruction on experimental data. On synthetic data, FlexSIM estimates the light pattern from the raw frames and uses an analytical transfer function at the nominal NA, wavelength, and pixel size. Sparse-SIM is applied to the conventional-SIM output in both cases. For each benchmark, its parameters are selected by visual inspection of reconstructions from separate calibration stacks and then held fixed across all evaluated stacks in that benchmark. No emitter reference enters this choice. The learned references are DFCAN, rDL-SIM, ML-SIM, and SSR-SIM, applied with their released weights and without target updates (with an additional author-trained BioSR clathrin-coated-pit DFCAN model in Appendix~\ref{app:baselines-detail}). Acquisition and baseline settings are given in Appendices~\ref{app:adapt-settings} and~\ref{app:baselines-detail}.

All synthetic outputs are resampled to a common threefold grid over matched physical fields of view. For the normalized root-mean-square error (NRMSE), both images are smoothed with a Gaussian kernel of \(\sigma=1.5\) output pixels (reflecting boundary), normalized independently by their 1st and 99.8th percentiles, and aligned by a least-squares scale and offset; for MS-SSIM, the same percentile normalization and clipped affine alignment are applied without pre-smoothing. Scores are averaged over the ten test stacks in each domain, using the final reconstruction at the fixed adaptation budget. Variants compared on the same stacks are evaluated with paired statistics: the mean paired difference with its \(95\%\) bootstrap interval (\(10{,}000\) resamples of stacks, stratified by domain when both domains are pooled) and the two-sided Wilcoxon signed-rank test. Appendix~\ref{app:metrics-detail} specifies the metrics and boundary handling.

Experimental comparisons use matched fields of view and profile coordinates. On the Argolight calibration target, peaks along fixed profiles across line pairs of known nominal spacing are counted with fixed criteria. Decorrelation analysis~\cite{descloux2019parameter} estimates the image decorrelation length scale within each output grid. The analysis procedures are given in Appendices~\ref{app:decorr-detail} and~\ref{app:expcompare-detail}.

\section{Results}

\begin{figure*}[!t]
    \centering
    \includegraphics[width=\textwidth]{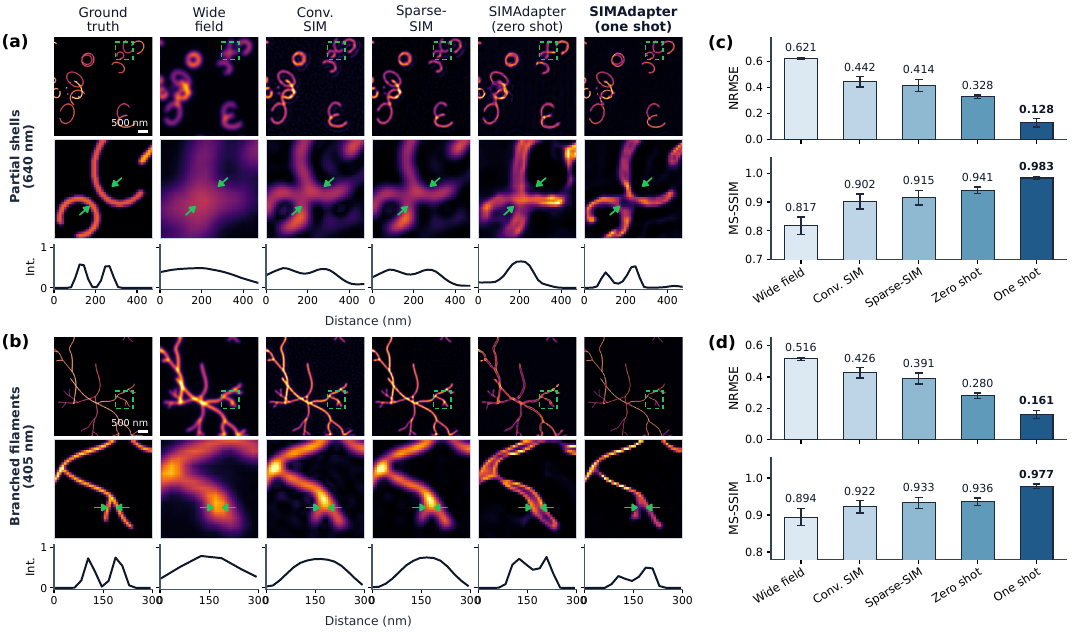}
    \caption{
        \textbf{Reconstruction accuracy across synthetic domain shifts.} (a,b) Ground-truth emitters and representative reconstructions for partial shells at \(640~\mathrm{nm}\) and branched filaments at \(405~\mathrm{nm}\), respectively. Comparisons include wide-field, conventional SIM, Sparse-SIM, the pretrained prior without adaptation (zero shot), and SIMAdapter after 1200 adaptation updates (one shot). Enlarged regions and profile coordinates are matched across methods. (c,d) Mean emitter NRMSE and MS-SSIM over \(10\) test stacks per domain; error bars indicate one standard deviation and labels give the means. Metrics follow the preprocessing and definitions in Section~\ref{sec:metrics}. Conventional SIM receives the simulated carrier geometry, modulation depth, and transfer function; Sparse-SIM parameters are chosen on separate calibration stacks (Section~\ref{sec:baselines}). Emitter ground truth is used only for evaluation. Scale bars, \(500~\mathrm{nm}\).
    }
    \label{fig:synthetic_transfer}
\end{figure*}

\subsection{Reconstruction accuracy across synthetic domains}
\label{sec:synthetic_results}

We first tested whether SIMAdapter could reconstruct emitter structures accurately when wavelength and morphology differed from synthetic pretraining. The two target domains contain partial shells at \(640~\mathrm{nm}\) and branched filaments at \(405~\mathrm{nm}\), with ten independently generated stacks in each domain. Numerical aperture and pixel size match pretraining, while defocus and blur are selected from the pretraining grid.

The reconstructions recover the overall shell arrangement and filament paths, but differ in the separation of nearby structures (Fig.~\ref{fig:synthetic_transfer}a,b). In the marked shell region, conventional SIM and Sparse-SIM produce broad profile maxima with a shallow intervening valley. SIMAdapter produces more distinct peaks at the two boundaries represented in the emitter reference. The difference is also visible in the branched-filament region: the classical reconstructions produce a broad peak across the marked pair, whereas SIMAdapter separates the two traces present in the reference.

The full-field metrics also favor SIMAdapter (Fig.~\ref{fig:synthetic_transfer}c,d and Table~\ref{tab:synthetic}). After \(1200\) adaptation updates, SIMAdapter reached emitter NRMSE values of \(0.128\) and \(0.161\), compared with \(0.414\) and \(0.391\) for Sparse-SIM. The corresponding MS-SSIM values were \(0.983\) and \(0.977\), compared with \(0.915\) and \(0.933\). Conventional SIM had NRMSE values of \(0.442\) and \(0.426\); wide-field imaging gave \(0.621\) and \(0.516\). Across the two domains, the mean NRMSE was \(0.144\) for SIMAdapter and \(0.403\) for Sparse-SIM, a reduction of \(64\%\), and SIMAdapter had the lower error on all \(20\) stacks.

We also applied published learned SIM reconstructions to the same stacks, using their released weights without target updates (Table~\ref{tab:synthetic}): DFCAN trained on BioSR microtubules or F-actin~\cite{qiao2021evaluation}, rDL-SIM, whose released microtubule networks denoise the raw frames before reconstruction~\cite{qiao2023rationalized}, ML-SIM, trained on randomized synthetic SIM data~\cite{christensen2021ml}, and the self-supervised SSR-SIM~\cite{liu2025bio}. The most accurate of these, ML-SIM, reached a mean NRMSE of \(0.315\) and an MS-SSIM of \(0.950\). Adapted SIMAdapter reduced the NRMSE by \(0.170\) (95\% CI \([0.159, 0.181]\)) and had the lower error on all \(20\) stacks (\(p=1.9\times10^{-6}\)); the same held against every other method in Table~\ref{tab:synthetic}. Before adaptation, the pretrained network was comparable to ML-SIM (NRMSE \(0.304\) and \(0.315\), \(p=0.28\)). Appendix~\ref{app:baselines-detail} shows the reconstructions and gives the inputs, transfer functions, and output grids of all methods.

\begin{table}[t]
\centering
\caption{Emitter reconstruction accuracy on the two held-out synthetic domains. Values are means over ten stacks per domain; learned methods use released weights without target updates unless noted.}
\label{tab:synthetic}
\footnotesize
\setlength{\tabcolsep}{2.8pt}
\begin{tabular}{lcccc}
\hline
 & \multicolumn{2}{c}{Shells, 640~nm} & \multicolumn{2}{c}{Filaments, 405~nm} \\
Method & NRMSE & MS-SSIM & NRMSE & MS-SSIM \\
\hline
Wide field & 0.621 & 0.817 & 0.516 & 0.894 \\
Conventional SIM & 0.442 & 0.902 & 0.426 & 0.922 \\
FlexSIM & 0.253 & 0.952 & 0.412 & 0.928 \\
Sparse-SIM & 0.414 & 0.915 & 0.391 & 0.933 \\
DFCAN (microtubules) & 0.509 & 0.887 & 0.896 & 0.663 \\
DFCAN (F-actin) & 0.412 & 0.919 & 0.365 & 0.942 \\
rDL-SIM (FlexSIM) & 0.336 & 0.940 & 0.563 & 0.886 \\
ML-SIM & 0.316 & 0.946 & 0.314 & 0.955 \\
SSR-SIM & 0.506 & 0.882 & 0.463 & 0.917 \\
SIMAdapter, zero shot & 0.328 & 0.941 & 0.280 & 0.936 \\
SIMAdapter, 1200 updates & \textbf{0.128} & \textbf{0.983} & \textbf{0.161} & \textbf{0.977} \\
\hline
\end{tabular}
\end{table}

\subsection{Contributions of pretraining and target adaptation}
\label{sec:adaptation_results}

We separated the contributions of pretraining and of the components of target adaptation with controlled comparisons on the same \(20\) synthetic stacks (Table~\ref{tab:ablation}). Every variant used the same raw frames, calibrated light-pattern anchor, bounded residual, loss, and budget of \(1200\) updates, and differed from the reference in a single factor. Differences are paired by stack and reported with \(95\%\) bootstrap intervals (\(10{,}000\) resamples, stratified by domain) and two-sided Wilcoxon signed-rank tests (Appendix~\ref{app:metrics-detail}).

Adaptation reduced the mean NRMSE of the mixed-condition network from \(0.304\) to \(0.144\), by \(0.160\) (\([0.150, 0.168]\)), and lowered the error on all \(20\) stacks; the mean MS-SSIM rose from \(0.938\) to \(0.980\). At the marked shell boundaries, the zero-shot profile shows one broad peak where the emitter reference has two, and adaptation separates them; at the marked filament pair, adaptation deepens the valley between the two traces (Fig.~\ref{fig:synthetic_transfer}a,b).

Pretraining accounts for much of the final accuracy. Started from randomly initialized weights, the same adaptation reached a mean NRMSE of \(0.528\) after \(1200\) updates, \(0.383\) (\([0.369, 0.398]\)) higher than from the mixed-condition network, and was less accurate on all \(20\) stacks (Holm-adjusted \(p=5.7\times10^{-6}\)). This variant received the same calibrated light pattern, trainable PSF, and number of updates, so these components alone do not reproduce the accuracy of the pretrained network.

Mixed-condition pretraining also improved on the BioSR-only network of our previous work, which initializes pretraining (Section~\ref{sec:corpus}). Before adaptation, the two networks were close for partial shells (NRMSE \(0.328\) and \(0.330\)), whereas the mixed-condition network was markedly more accurate for branched filaments (\(0.280\) and \(0.683\)). After adaptation, the mixed-condition network had the lower mean NRMSE, \(0.144\) against \(0.172\) (difference \(0.028\), \([0.011, 0.052]\), \(p=0.0014\)), with lower error on \(18\) of \(20\) stacks, and the higher MS-SSIM (difference \(0.013\), \(p=3.8\times10^{-6}\)). The NRMSE difference favored the mixed-condition network on all ten filament stacks (\(p=0.002\)); for partial shells its interval included zero (\([-0.003, 0.075]\), \(p=0.19\)), while the MS-SSIM difference favored the mixed-condition network (\(p=0.002\)). Confidence intervals reflect paired variation across the ten test stacks per domain.

Among the components of adaptation, PSF optimization had the largest effect. Holding the initial PSF predicted before adaptation fixed raised the mean NRMSE to \(0.277\), by \(0.133\) (\([0.124, 0.142]\)), on all \(20\) stacks. Freezing the shared masked-autoencoder backbone, \(78.2\%\) of the parameters, and updating only the emitter readout, low-rank, and physics branches produced no detectable change in NRMSE (\(0.144\); difference \(-0.001\), \([-0.008, 0.006]\), \(p=0.25\)). Removing the dark-region penalty changed the NRMSE by \(-0.002\) (\([-0.005, 0.000]\), \(p=0.13\)); its effect on experimental line pairs is examined in Section~\ref{sec:experimental_results}. The synthetic reconstructions of these variants (Appendix~\ref{app:adapt-settings}) show that random initialization fails to resolve fine structures and that fixing the PSF blurs shell boundaries and filaments. On the experimental Argolight target, random initialization separated none of the central double lines (\(0.0\%\)) and kept \(0.0\%\) of the flanking single lines single-peaked, holding the PSF fixed separated the nominal \(60~\mathrm{nm}\) double line in \(44.6\%\) of rows, and freezing the backbone separated it in \(85.9\%\), against \(83.5\%\) for the reference, with \(100.0\%\) of the single lines kept single by the latter three variants (Appendix~\ref{app:adapt-settings}).

\begin{table*}[t]
\centering
\caption{Controlled comparisons at \(1200\) updates on the \(20\) synthetic stacks. Each variant differs from the reference in one factor. \(\Delta\) is the paired mean NRMSE difference from the reference with its \(95\%\) bootstrap interval.}
\label{tab:ablation}
\footnotesize
\setlength{\tabcolsep}{6pt}
\begin{tabular}{lcccc}
\hline
 & Shells & Filaments & \multicolumn{2}{c}{Both domains} \\
Variant & NRMSE & NRMSE & NRMSE & \(\Delta\) [95\% CI] \\
\hline
Reference: mixed pretraining & 0.128 & 0.161 & 0.144 & -- \\
BioSR-only pretraining & 0.156 & 0.189 & 0.172 & \(+0.028\) [\(0.011, 0.052\)] \\
Random initialization & 0.533 & 0.522 & 0.528 & \(+0.383\) [\(0.369, 0.398\)] \\
Fixed initial PSF & 0.289 & 0.266 & 0.277 & \(+0.133\) [\(0.124, 0.142\)] \\
Backbone frozen & 0.126 & 0.161 & 0.144 & \(-0.001\) [\(-0.008, 0.006\)] \\
No dark-region penalty & 0.125 & 0.160 & 0.142 & \(-0.002\) [\(-0.005, 0.000\)] \\
\hline
\end{tabular}
\end{table*}

\subsection{Unseen optical conditions}
\label{sec:optics_results}

The two held-out domains of Section~\ref{sec:synthetic_results} keep the numerical aperture, pixel size, and illumination period of synthetic pretraining. We therefore repeated the test on the same \(20\) emitters under optical conditions outside the pretraining grid, with the expected photon count of each stack matched to its reference (Table~\ref{tab:optics}). Two pupil aberrations, horizontal astigmatism and horizontal coma at \(0.10\) waves RMS (Strehl ratio \(0.67\)), exceed the Mar\'echal criterion for diffraction-limited imaging. Three simulated configurations change the PSF scale in detector pixels: NA \(1.49\) with \(65~\mathrm{nm}\) pixels (H1), NA \(1.10\) with \(55~\mathrm{nm}\) pixels and a seven-pixel illumination period (H2), and H1 with \(0.060\) waves RMS horizontal astigmatism and \(0.053\) waves RMS horizontal coma (\(0.08\) waves RMS combined; H3). The product of NA and pixel size, which sets the PSF width in pixels at a given wavelength, spans \(60.5\)--\(96.9~\mathrm{nm}\), against \(81.4\)--\(91.2~\mathrm{nm}\) in pretraining (Appendix~\ref{app:corpus}). Each stack was calibrated from its own raw frames and adapted for \(1200\) updates with the PSF token of its domain.

Adaptation reduced reconstruction error across the tested \(0.10\)-wave aberrations and H1--H3 configurations (Table~\ref{tab:optics}). In the three simulated configurations, the mean NRMSE over both domains fell from \(0.265\) to \(0.130\) (H1), from \(0.471\) to \(0.215\) (H2), and from \(0.304\) to \(0.163\) (H3), with lower error on all \(20\) stacks in each case (\(p=1.9\times10^{-6}\); Appendix~\ref{app:additional}). It remained below that of FlexSIM (\(0.355\)--\(0.364\)) and Wiener-SIM (\(0.529\)--\(0.639\)) reconstructions calibrated from the same raw frames. Under astigmatism, the NRMSE fell from \(0.445\) to \(0.228\) for shells and from \(0.295\) to \(0.157\) for filaments. Under coma, adaptation reduced the filament NRMSE from \(0.370\) to \(0.182\). For partial shells, joint adaptation introduced a \(28.8\pm2.5~\mathrm{nm}\) (\(0.46\) raw pixel) lateral translation along the coma axis, consistent with centering an asymmetric coma PSF, which left pixel-locked NRMSE unchanged at \(0.433\) (\(0.433\rightarrow0.433\)); under matched rigid shift registration at both endpoints (each reconstruction independently aligned to the ground truth by subpixel phase correlation with upsampling factor \(50\)), NRMSE fell from \(0.497\) to \(0.250\) and MS-SSIM rose from \(0.865\) to \(0.938\) (Appendix~\ref{app:additional}). At twice this aberration strength (\(0.20\) waves RMS, diffraction Strehl ratio \(0.20\)--\(0.22\)), adaptation still improved filament astigmatism (NRMSE \(0.698\) to \(0.306\), MS-SSIM \(0.801\) to \(0.956\)), left shell astigmatism nearly unchanged (\(0.785\) to \(0.783\)), and increased the error under \(0.20\)-wave coma (NRMSE \(0.787\) to \(0.896\) for shells and \(0.816\) to \(0.844\) for filaments; pooled MS-SSIM \(0.677\) to \(0.610\)).

\begin{table}[t]
\centering
\caption{Reconstruction under unseen optical conditions. Mean emitter NRMSE over ten stacks per domain before and after \(1200\) adaptation updates. The last column gives the mean NRMSE over both domains for FlexSIM calibrated from the same raw frames. H1--H3 are simulated configurations; \(\Delta x\) is detector pixel size.}
\label{tab:optics}
\scriptsize
\setlength{\tabcolsep}{2.4pt}
\begin{tabular}{lccc}
\hline
Condition & Shells & Filaments & FlexSIM \\
\hline
Pretraining (NA 1.30, 62.6~nm) & \(0.328\!\rightarrow\!0.128\) & \(0.280\!\rightarrow\!0.161\) & 0.333 \\
Astigmatism, 0.10 waves & \(0.445\!\rightarrow\!0.228\) & \(0.295\!\rightarrow\!0.157\) & -- \\
Coma, 0.10 waves & \(0.497\!\rightarrow\!0.250^{a}\) & \(0.370\!\rightarrow\!0.182\) & -- \\
H1: NA 1.49, \(\Delta x\) 65~nm & \(0.255\!\rightarrow\!0.100\) & \(0.275\!\rightarrow\!0.161\) & 0.355 \\
H2: NA 1.10, 55~nm, 7-px & \(0.650\!\rightarrow\!0.258\) & \(0.292\!\rightarrow\!0.172\) & 0.364 \\
H3: H1 + astig.\ \& coma\(^{b}\) & \(0.317\!\rightarrow\!0.156\) & \(0.292\!\rightarrow\!0.171\) & 0.364 \\
\hline
\end{tabular}
\par\smallskip{\scriptsize\raggedright \(^{a}\)Each reconstruction independently aligned to the ground truth by subpixel phase correlation (upsampling factor \(50\)) before scoring (pixel-locked NRMSE is \(0.433\!\rightarrow\!0.433\) owing to the \(28.8\pm2.5~\mathrm{nm}\) lateral translation along the coma axis). \(^{b}\)\(0.060\) waves RMS horiz.\ astigmatism and \(0.053\) waves RMS horiz.\ coma (\(0.08\) waves RMS combined); H2 uses a 7-pixel illumination period.\par}
\end{table}

Because the synthetic stacks were generated with known kernels, the learned PSF can be compared with the true one (Appendix~\ref{app:additional}). The kernel predicted from a held-out token before adaptation has several lobes. After adaptation, its centered relative \(\ell_2\) distance to the true kernel fell from \(0.447\) to \(0.084\) for shells and from \(0.420\) to \(0.214\) for filaments, on all \(20\) stacks. For shells, the width and radial MTF of the learned kernel closely match those of the true kernel, with the FWHM error falling from \(18.2\) / \(52.2~\mathrm{nm}\) to \(5.9\) / \(4.2~\mathrm{nm}\) along \(x\) and \(y\). For filaments, the learned kernel is \(7\%\) wider at half maximum than the true kernel (\(172\) against \(160~\mathrm{nm}\)) and exhibits faster mid-frequency roll-off in its radial MTF, while the emitter NRMSE still fell from \(0.280\) to \(0.161\). Joint estimation therefore brings the centered kernel shape close to the truth but leaves a residual trade-off between the frequency response of the PSF and that of the emitter. In H1--H3, the centered distance fell from \(0.450\) to \(0.184\), from \(0.616\) to \(0.143\), and from \(0.570\) to \(0.288\).

The light pattern of each stack is calibrated from its raw frames, and the bounded residual of Section~\ref{sec:recon} corrects small calibration errors during adaptation. To test the sensitivity to calibration errors, we shifted the phase (\(\pm10^\circ\), \(\pm30^\circ\)) or the carrier frequency (\(\pm0.5\%\), \(\pm2\%\)) of one orientation of the calibrated pattern of each synthetic stack (Appendix~\ref{app:additional}). For \(\pm10^\circ\) and \(\pm0.5\%\), adaptation reduced the pooled signed offsets relative to the unperturbed \(1200\)-update control to \(-2.26^\circ/+2.15^\circ\) and \(-0.217\%/+0.225\%\), and the absolute pooled mean NRMSE and MS-SSIM changes remained at most \(0.0044\) and \(0.0012\). A~\(+30^\circ\)~phase error raised the NRMSE by \(0.012\), and frequency errors of \(\pm2\%\), which exceed the \(\pm1\%\) range of the residual, raised it by \(0.032\) and \(0.034\). For comparison, the median errors of the unperturbed calibration were \(0.3\)--\(1.8^\circ\) in phase and \(0.05\)--\(0.15\%\) in frequency.

\begin{figure*}[!t]
    \centering
    \includegraphics[width=0.75\textwidth]{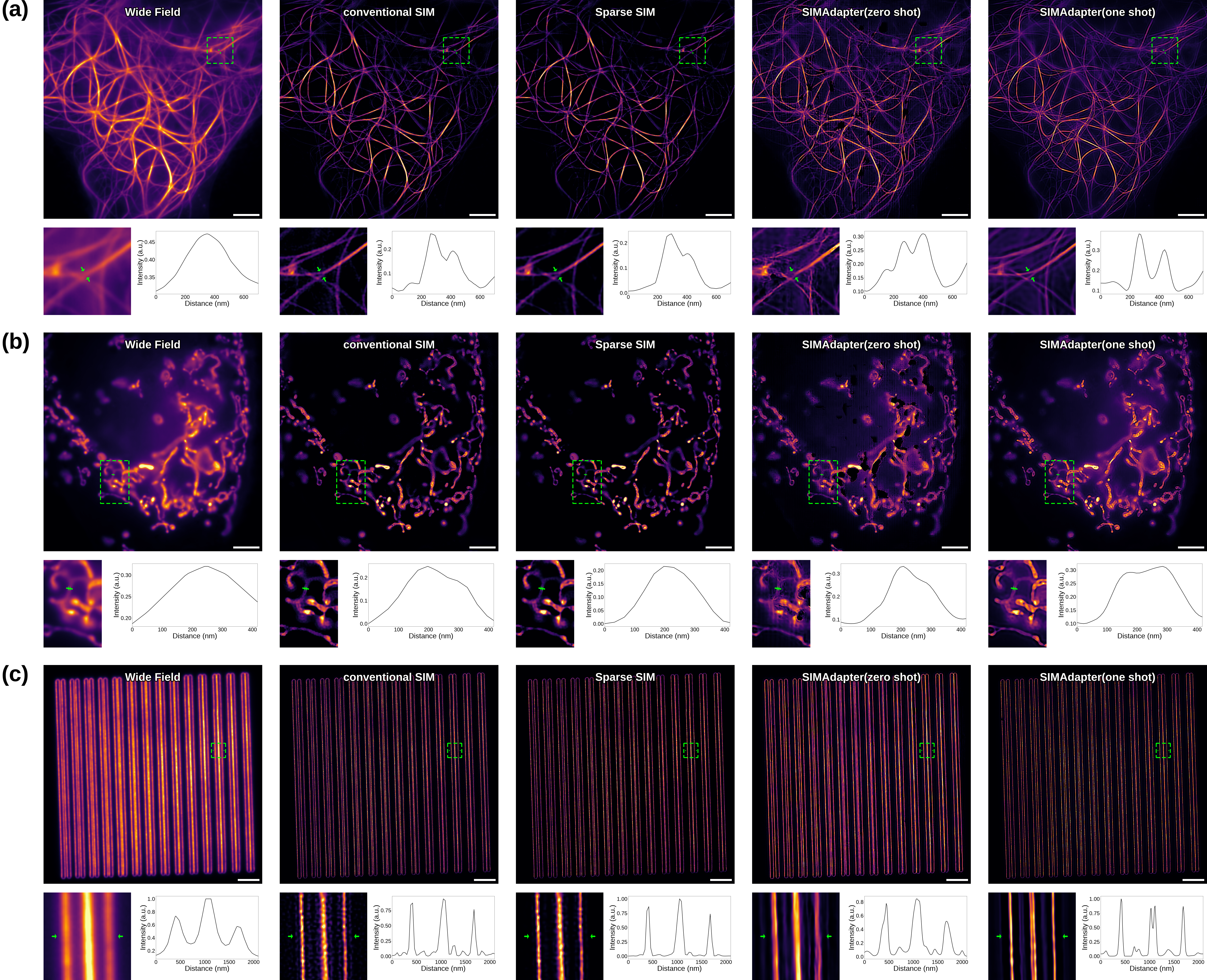}
    \caption{
        \textbf{Reconstructions of external measurements.} (a) FlexSIM Stand microtubules. (b) FlexSIM Stand mitochondria. (c) FlexSIM Stand Argolight, a calibration target with line pairs of known nominal spacing. Columns show wide field, conventional SIM, Sparse-SIM, and SIMAdapter before (zero shot) and after (one shot) 1200 adaptation updates. Each scene contributes one stack excluded from pretraining and used for both unlabeled adaptation and reconstruction. Dashed boxes mark the enlarged regions, and arrows mark the profile endpoints, matched across methods. The profiles cross a filament pair in (a), mitochondrial boundaries in (b), and the nominal \(60~\mathrm{nm}\) group in (c), a central double line between two single reference lines. Each image is displayed between its full-field 0.1st and 99.9th percentiles, and each profile has its own vertical axis. Scale bars, \(4~\mu\mathrm{m}\).
    }
    \label{fig:external_transfer}
\end{figure*}

\subsection{Local structure in external measurements}
\label{sec:experimental_results}

\begin{figure*}[!t]
    \centering
    \includegraphics[
        width=0.80\textwidth
    ]{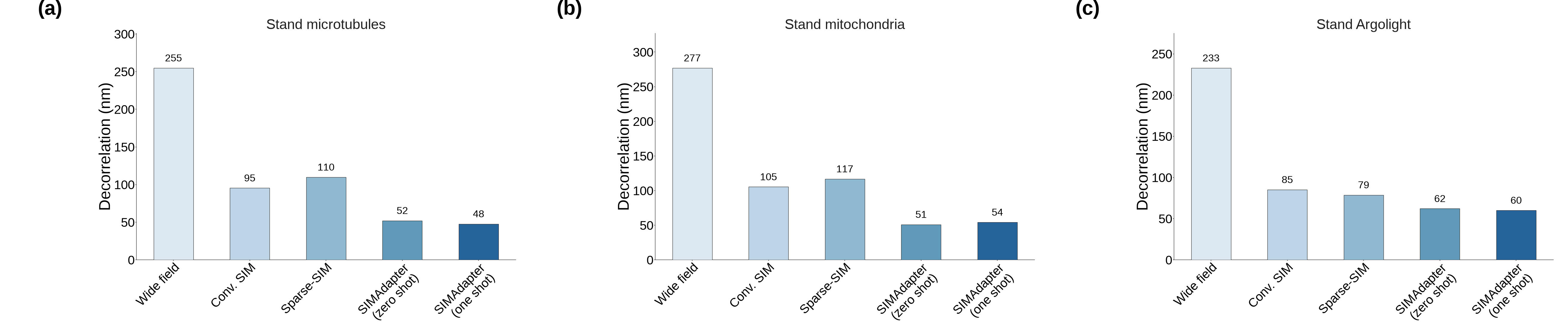}
    \caption{
        \textbf{Image decorrelation length scales of the experimental reconstructions.}
        (a) Stand microtubules, (b) Stand mitochondria, and (c) Stand
        Argolight. Each bar gives the full-field decorrelation length-scale estimate
        of the corresponding reconstruction in
        Fig.~\ref{fig:external_transfer}, computed on the native output grid of each method: \(1\times\) for wide field, \(2\times\) for conventional SIM and Sparse-SIM, and \(3\times\) for SIMAdapter. The estimates depend on image content and sampling.
    }
    \label{fig:decorrelation}
\end{figure*}

Figure~\ref{fig:external_transfer} compares matched regions and intensity profiles from the three experimental stacks, each adapted with the same budget of 1200 updates, alongside the classical reconstructions.

The Stand microtubules stack contains an extended network of closely spaced and crossing filaments (Fig.~\ref{fig:external_transfer}a). Wide-field imaging renders the marked pair as one broad feature. Conventional SIM and Sparse-SIM narrow the traces, but their profiles retain substantial intensity between the dominant and secondary maxima. The zero-shot reconstruction contains dark voids and a grid-like texture across the field, and its profile shows several maxima. After adaptation, the voids and texture are absent and the profile shows two distinct maxima separated by a deep valley.

At the marked mitochondrial boundaries (Fig.~\ref{fig:external_transfer}b), wide-field imaging, conventional SIM, and Sparse-SIM give broad profiles with one maximum. The zero-shot reconstruction again contains dark voids, and its profile shows one maximum with a shoulder. The adapted reconstruction shows continuous boundary traces, and its profile has a maximum at each boundary with a shallow dip between them.

Stand Argolight is a calibration target in which each line group places a central double line between two single reference lines, a single--double--single structure of known nominal spacing (Fig.~\ref{fig:external_transfer}c). Because single and double lines appear side by side under identical imaging conditions, each group tests whether a reconstruction separates two lines without splitting one. Figure~\ref{fig:external_transfer}c shows the cross-section of the nominal \(60~\mathrm{nm}\) group: wide-field imaging, conventional SIM, Sparse-SIM, and zero-shot SIMAdapter each show one peak at each single line and one peak at the central double line, whereas adapted SIMAdapter shows one peak at each single line and two peaks at the central double line (spacing \(78.6~\mathrm{nm}\), minimum-to-peak ratio \(0.565\)). Across \(1200\) rows (\(y=200\text{--}1399\)) of the nominal \(120\), \(90\), and \(60~\mathrm{nm}\) groups (Appendix~\ref{app:adapt-settings}), adapted SIMAdapter kept \(100\%\) of the flanking single lines single-peaked and separated the central double line in \(100.0\%\), \(99.8\%\), and \(83.5\%\) of rows (median peak spacings \(131.0\), \(104.8\), and \(78.6~\mathrm{nm}\); median minimum-to-peak ratios \(0.137\), \(0.408\), and \(0.626\)). The \(60~\mathrm{nm}\) spacing lies below the linear-SIM resolution limit of about \(93~\mathrm{nm}\) (\(\lambda/4\mathrm{NA}\) for NA \(1.42\) and \(527~\mathrm{nm}\) emission); its separation therefore reflects extrapolation beyond the SIM passband by the learned prior, while the single lines in the same rows remain single. By contrast, wide-field imaging separates none of the three double lines, zero-shot SIMAdapter separates at most \(5.1\%\) while splitting \(21\text{--}35\%\) of the single lines, conventional SIM separates \(96.5\%\) at \(120~\mathrm{nm}\) and \(42.3\%\) at \(90~\mathrm{nm}\) with sidelobe peaks on about half of the single lines, and Sparse-SIM separates \(93.6\%\) at \(120~\mathrm{nm}\) (\(0.375\)), \(26.2\%\) at \(90~\mathrm{nm}\) (\(0.686\)), and none at \(60~\mathrm{nm}\).

The adaptation objective includes a penalty on emitter intensity in dark measurement regions (Section~\ref{sec:adapt}). Setting its weight \(\lambda_d\) to zero, with all other settings unchanged, leaves both the single- versus double-line distinction and all three Argolight regions essentially unchanged (Appendix~\ref{app:adapt-settings}): flanking single reference lines remain \(100\%\) single-peaked, and the two-peak fraction and median minimum-to-peak ratio change only from \(100.0\% / 0.137\) to \(100.0\% / 0.149\) at \(120~\mathrm{nm}\), from \(99.8\% / 0.408\) to \(97.8\% / 0.418\) at \(90~\mathrm{nm}\), and from \(83.5\% / 0.626\) to \(82.2\% / 0.634\) at \(60~\mathrm{nm}\). Weak traces in the microtubule and mitochondria scenes also persist with and without the penalty (Appendix~\ref{app:adapt-settings}), and on the \(20\) synthetic stacks removing the penalty changes the mean NRMSE by \(-0.002\) (Table~\ref{tab:ablation}). The separation of the line pairs therefore does not arise from the dark-region penalty.

We also tested whether a network adapted to one stack reconstructs another stack acquired under the same conditions without further updates. Two separately recorded nine-frame mitochondria stacks from the PCA-SIM microscope~\cite{qian2023structured} (NA \(1.40\), \(488~\mathrm{nm}\) excitation, \(65~\mathrm{nm}\) pixels), absent from the pretraining corpus, were each adapted for \(1200\) updates and then used as the source for the other. With no update on the target, the transferred network reduced the relative residual between the reprojected and measured frames from \(0.257\) to \(0.110\) for stack B and from \(0.241\) to \(0.133\) for stack A, close to the residuals of the networks adapted to the targets themselves (\(0.109\) and \(0.135\); Appendix~\ref{app:additional}). The transferred and self-adapted emitters differed by \(3.3\)--\(3.6\%\) in relative \(\ell_2\) norm, against \(30\)--\(32\%\) between the zero-shot and self-adapted emitters.
On the synthetic domains, which have emitter references, we adapted the network to the first stack of each domain and applied it without updates to the other nine stacks. The transferred network reduced the mean NRMSE from \(0.305\) without adaptation to \(0.187\), with lower error on all \(18\) stacks (\(p=7.6\times10^{-6}\)), against \(0.145\) for adaptation to each stack itself. It thus recovered \(74\%\) of the gain of per-stack adaptation, \(65\%\) for partial shells and \(88\%\) for branched filaments (Appendix~\ref{app:additional}).

\subsection{Image decorrelation length scales}
\label{sec:resolution_results}

Figure~\ref{fig:decorrelation} reports the decorrelation length scale~\cite{descloux2019parameter} of each experimental reconstruction. This image statistic summarizes the spatial-frequency content of a reconstruction and depends on its content and sampling grid, so we use it to compare reconstructions of the same acquisition; the line-pair measurement of Section~\ref{sec:experimental_results} tests separation against a known structure. Estimates near the two-pixel limit of the threefold grid (\(43.3~\mathrm{nm}\) for microtubules and mitochondria, \(52.4~\mathrm{nm}\) for Argolight) reflect the output sampling and do not measure optical resolution. After adaptation, the full-field estimate changed from \(52.2\) to \(47.8~\mathrm{nm}\) for Stand microtubules, from \(51.1\) to \(54.3~\mathrm{nm}\) for Stand mitochondria, and from \(62.2\) to \(60.1~\mathrm{nm}\) for Stand Argolight. Across five fixed subfields of each adapted reconstruction, the estimates ranged from \(44.8\) to \(49.8~\mathrm{nm}\), \(44.3\) to \(52.2~\mathrm{nm}\), and \(52.6\) to \(62.1~\mathrm{nm}\), respectively (Appendix~\ref{app:decorr-detail}).

\section{Discussion}

SIMAdapter uses a shared pretrained network to initialize measurement-specific reconstruction, and each raw stack then constrains the emitter and optical model through the SIM forward model. The controlled comparisons show that both stages contribute. Adaptation from random weights, with the same calibrated light pattern, PSF optimization, and number of updates, remained far less accurate, and adaptation from the mixed-condition network was more accurate than adaptation from the BioSR-only network of our previous work. Among the components of adaptation, refining the PSF had the largest effect, whereas freezing the shared backbone produced no detectable loss in accuracy.

The prior also transferred to optical conditions outside the pretraining grid. Across the tested \(0.10\)-wave aberrations and simulated changes of numerical aperture, pixel size, and illumination period (H1--H3), adaptation reduced reconstruction error and brought the centered shape of the learned PSF close to the true kernel, starting from an initial kernel set by a held-out condition token. During pretraining, masked-autoencoder reconstruction through the forward convolution constrains PSF learning directly from microscopy observations, as previously demonstrated for physics-informed masked autoencoders~\cite{wang2024learning}; during single-stack adaptation, the calibrated light pattern, PSF regularization, and pretrained emitter prior further constrain the joint solution. The same tests show where joint estimation of emitter and PSF remains approximate: the filament kernel converged to a centered shape close to the truth but remained \(7\%\) wider at half maximum with residual mid-frequency mismatch, under \(0.10\)-wave coma the shell reconstruction was displaced by about half a raw pixel, and stronger \(0.20\)-wave coma exceeded the recovery range of single-stack adaptation. A physical parameterization of the PSF, upsampled PSF modeling~\cite{chen2025upsampled}, or a measured PSF where one is available, could constrain these cases further.

In the experimental scenes, adaptation removed the voids and texture of the direct predictions, and on the Argolight target it distinguished flanking single reference lines from central double line pairs across the tested nominal \(120\), \(90\), and \(60~\mathrm{nm}\) spacings, with or without the dark-region penalty. The biological scenes have no emitter reference, so their profiles describe local structure; the synthetic domains and the calibration target provide the quantitative tests.

With the fixed budget of \(1200\) updates, adaptation takes a median of \(5.9\) minutes per stack on one NVIDIA A800-SXM4-40GB GPU. Most of the synthetic error reduction occurs within the first \(250\)--\(500\) updates (Appendix~\ref{app:adapt-settings}), and a network adapted to one stack reconstructed other stacks acquired under the same conditions without further updates, recovering about three quarters of the gain of per-stack adaptation on the synthetic domains (Section~\ref{sec:experimental_results}). Both observations point to lower costs for image series. Extending the approach to other imaging modalities would require an appropriate forward model for each modality.

\section{Conclusion}

SIMAdapter combines a self-supervised SIM reconstruction prior with adaptation to a single unlabeled raw stack. In the two held-out synthetic domains, the adapted reconstructions were substantially more accurate than the pretrained model's direct predictions, the classical references, and published learned reconstructions, and both pretraining and adaptation contributed to this accuracy. Adaptation also reduced reconstruction error across the tested \(0.10\)-wave aberrations and pixel-space PSF scales, and on an experimental calibration target it distinguished flanking single reference lines from central double line pairs across the tested nominal \(120\), \(90\), and \(60~\mathrm{nm}\) spacings. A common pretrained network can therefore be reused for reconstruction of individual SIM stacks without paired high-resolution training data.


\appendix
\renewcommand{\thesubsection}{\thesection.\arabic{subsection}}
\makeatletter
\renewcommand{\p@subsection}{}
\@addtoreset{equation}{section}
\makeatother
\renewcommand{\theequation}{\thesection\arabic{equation}}
\titleformat{\section}
  {\large\sffamily\bfseries\filright} {APPENDIX \thesection:} {0.5em} {\MakeUppercase{#1}} []

\section{Light-pattern calibration and normalization}

\subsection{Light-pattern normalization conventions and the bounded residual}\label{app:normalization}
The light-pattern calibration \(\mathcal{C}\) (Appendix~\ref{app:calibration}) returns a solver-normalized anchor \(\bar{\mathbf{P}}_i=\mathcal{C}[\mathcal{B}(\mathbf{x}_i)]\) with joint stack mean \(1/T\) across its \(T=9\) frames (and per-frame unit-mean normalization applied before cropping during adaptation). Multiplying by \(T\) gives unit-mean scaling before applying the bounded residual. The residual has \(24\) parameters: nine gains, nine phases, and three two-dimensional wave vectors:
\begin{equation}
    \widehat{\mathbf{P}}_i
    =
    \mathcal{A}
    \left(
        \bar{\mathbf{P}}_i,
        \boldsymbol{\delta}_i
    \right),
    \label{eq:illumination_residual}
\end{equation}
where \(\mathcal{A}\) applies these adjustments at the sample centers of the threefold grid in detector coordinates, the convention used to render the anchor. The gains are parameterized as \(T\) times a softmax over bounded logits, so their sum is exactly \(T\). This fixes the mean frame gain at one without renormalizing each adjusted frame. Individual frame means can therefore differ from their nominal values: the factor \(T\) in \eqref{eq:sim_forward_frame} converts the anchor convention rather than setting every adjusted frame to unit mean.

\subsection{FlexSIM preprocessing and light-pattern calibration}\label{app:calibration}
We use the published FlexSIM preprocessing operators~\cite{soubies2024surpassing}, with the constants and rejection rule specified below. Background correction \(\mathcal{B}\) subtracts the mean of a square low-signal region from each frame and clips the result at zero. The region's side length is fixed for each acquisition family and ranges from 15 to 257 detector pixels across the 23 families. Stacks from families without a specified background region are unchanged.

Calibration \(\mathcal{C}\) suppresses the wide-field component and estimates one carrier wave vector and one phase per orientation from the background-corrected stack. Crop size, wide-field stop radius, search annulus, and modulation depth are fixed for each acquisition family. Modulation depths range from \(0.5\) to \(1.0\), and wave vectors are expressed in radians per nanometre. The experimental scene names follow the FlexSIM acquisition families. For Stand microtubules, Stand mitochondria, and Stand Argolight, the background-region side lengths are \(101\), \(131\), and \(31\) pixels, respectively, and the modulation depths are \(0.8\), \(0.6\), and \(0.6\). All three use the full frame for calibration and the search annulus \([0,1]\). Constants for all 23 families are available from the authors on request.

For BioSR and BioTISR, the carrier search is restricted to a band around the nominal illumination frequency of each acquisition family, and the three orientations are constrained to equal carrier magnitude. When the carrier estimate of a stack is unstable, its wave vectors are replaced by the consensus of its acquisition group or, for BioTISR lysosomes and mitochondria, by the consensus of training cells of the same structure; the phases are always estimated from the stack itself. On \(45\) internal BioTISR training-split confirmation stacks (three cells per structure excluded from estimating the fallback consensus), all \(45\) final anchors agreed with the leave-one-out consensus within \(1\%\) in frequency and \(1^\circ\) in orientation.

We render the calibrated light pattern analytically from the estimated wave vectors and phases on the threefold-upsampled grid, at the sample centers in detector coordinates. An optional low-frequency envelope correction (set uniform for the executed mixed pretraining and Stand target runs) and the normalization of Appendix~\ref{app:normalization} are applied. A stack is excluded from pretraining if calibration fails to return three finite wave vectors and three finite phases, if the pattern normalizer is non-positive, or if any wave-vector norm is below \(10^{-6}\) radians per nanometre, which indicates a collapsed carrier. An experimental stack is also excluded if its carriers do not belong to a physically consistent group shared by at least \(60\%\) of the stacks of its acquisition group; this rule excludes \(55\) stacks.

\section{Scenario-indexed PSF conditioning}\label{app:psfcond}
The PSF branch, shared across conditions, maps each condition token to an effective kernel:
\begin{equation}
    \mathbf{h}_{s_i}
    =
    \operatorname{softmax}
    \left[
        g_{\psi}
        \left(
            \mathbf{t}_{s_i}
        \right)
    \right],
    \label{eq:psf_prediction}
\end{equation}
where \(g_\psi\) is the convolutional PSF branch and \(\mathbf{t}_{s_i}\) is the token of condition \(s_i\). Spatial softmax makes the kernel non-negative and normalizes its sum to one. Each raw SIM stack used during pretraining carries a source-level optical-condition index \(s_i\in\mathcal{S}_{\mathrm{pre}}\), where \(\mathcal{S}_{\mathrm{pre}}\) is the set of conditions represented in the pretraining corpus. The tokens form a fixed table of \(32\) entries with the shape of the low-rank PSF representation. Each entry is a shared base code plus a deterministic pseudo-random code generated from the condition name; the entries are set before training and are never optimized. Fourteen entries occur in pretraining: one for BioSR, one for BioTISR, and twelve for the distinct synthetic kernels of Appendix~\ref{app:corpus}. Two separate reserved entries, absent from pretraining, are assigned to the held-out branched-filament and partial-shell domains.

The same emitter network and PSF branch are used for all conditions. The condition index determines the PSF condition token supplied to this model; it does not select a separately trained reconstruction model. Raw-domain self-supervision updates the model parameters while the tokens remain fixed. The light pattern is calibrated for each stack and is not encoded by the PSF condition token.

A token identifies a condition and encodes no optical parameter. The tokens let one shared emitter network and PSF branch fit condition-specific kernels during pretraining; for a test stack, the kernel of its token is the starting point that adaptation refines from the target frames. The kernel predicted from a reserved token before adaptation is therefore a fixed starting point, and the PSF of a target is determined by adaptation against its raw frames (Section~\ref{sec:optics_results}). The synthetic held-out domains use their reserved tokens. The experimental scenes, whose optics differ from every pretraining condition, use the BioSR entry, so their adaptation starts from the kernel learned for BioSR. The token remains fixed during adaptation.

\section{Complete self-supervised training objective}\label{app:objective-full}
The pretraining loss of \eqref{eq:training_loss} is averaged over each mini-batch, with \(\lambda_h=10^{-3}\), \(\lambda_P=10^{-3}\), \(\lambda_c=10^{-1}\), and \(\lambda_\delta=10^{-3}\). The raw-stack reconstruction term is
\begin{equation}
\begin{aligned}
    \mathcal{L}_{\mathrm{rec}}^{(i)}
    &=
    0.875
    \frac{
        \left\|
            \mathbf{M}_i
            \odot
            \left(
                \tilde{\mathbf{y}}_i-\mathbf{y}_i
            \right)
        \right\|_1
    }{
        \left\|
            \mathbf{M}_i
        \right\|_1
    }
    \\
    &\quad+
    0.125
    \left[
        1-
        \operatorname{MS\text{-}SSIM}
        \left(
            \tilde{\mathbf{y}}_i,
            \mathbf{y}_i
        \right)
    \right].
\end{aligned}
\label{eq:rec_loss}
\end{equation}
The masked-autoencoder reconstruction mask \(\mathbf{M}_i\) applies only to the pixel-wise \(\ell_1\) term; it does not select the foreground. The structural-similarity term evaluates 3D MS-SSIM (window size \(5\)) on the complete raw stack scaled by the observation minimum and maximum. With masked reconstruction disabled, \(\mathbf{M}_i\) is an all-one tensor. The PSF \(\mathbf{h}_{s_i}\) and the light pattern \(\widehat{\mathbf{P}}_i\), denoted generically by \(\mathbf{v}\), are regularized by the sum over represented non-singleton dimensions of the mean squared forward finite difference of their mean-normalized values:
\begin{equation}
\begin{aligned}
    \bar{\mathbf{v}}
    &=
    \frac{
        \mathbf{v}
    }{
        \operatorname{mean}
        \left(
            \mathbf{v}
        \right)
    },
    \\
    \mathcal{R}_{\mathrm{grad}}
    \left(
        \mathbf{v}
    \right)
    &=
    \sum_d
    \operatorname{mean}\!\left[
        \left(\nabla_d\bar{\mathbf{v}}\right)^2
    \right],
\end{aligned}
\label{eq:gradient_regularization}
\end{equation}
where \(\nabla_d\) is the forward finite difference along non-singleton tensor dimension \(d\) (the two spatial axes for the PSF and the frame plus two spatial axes for the nine-frame light pattern).

The PSF center penalty is
\begin{equation}
    \mathcal{R}_{\mathrm{center}}
    \left(
        \mathbf{h}
    \right)
    =
    \left\|
        \frac{
            \displaystyle
            \sum_{\mathbf{r}\in\Omega_h}
            \mathbf{r}\mathbf{h}(\mathbf{r})
        }{
            \displaystyle
            \sum_{\mathbf{r}\in\Omega_h}
            \mathbf{h}(\mathbf{r})
        }
        -
        \mathbf{c}_0
    \right\|_2,
    \label{eq:center_regularization}
\end{equation}
where \(\Omega_h\) is the PSF support, \(\mathbf{r}\) denotes coordinates normalized to \([-1,1]\), and \(\mathbf{c}_0\) is the center of the support.

The residual penalty is
\begin{equation}
\begin{aligned}
    \mathcal{R}_{\delta}
    \left(
        \boldsymbol{\delta}_i
    \right)
    ={}&
    \operatorname{mean}\!\left[\tanh^2(\boldsymbol{\delta}_{i,g})\right]
    +
    \operatorname{mean}\!\left[\tanh^2(\boldsymbol{\delta}_{i,\phi})\right]
    \\
    &+
    \operatorname{mean}\!\left[\tanh^2(\boldsymbol{\delta}_{i,k})\right],
\end{aligned}
    \label{eq:lp_residual_regularization}
\end{equation}
where \(\boldsymbol{\delta}_{i,g}\), \(\boldsymbol{\delta}_{i,\phi}\), and \(\boldsymbol{\delta}_{i,k}\) are the gain, phase, and wave-vector residuals, respectively.

\section{Pretraining corpus and protocol}
\subsection{Corpus composition and data splits}\label{app:corpus}
Table~\ref{tab:corpus} lists the corpus before training stacks with invalid light-pattern calibration are excluded. BioSR contains images of clathrin-coated pits, endoplasmic reticulum, F-actin, and microtubules~\cite{qiao2021evaluation}. BioTISR contains images of clathrin-coated pits, F-actin, lysosomes, microtubules, and mitochondria. We use its level-03, cell-disjoint splits~\cite{qiao2026neural} and exclude levels 01 and 02. A single source-level token identifies the dataset-level optical condition of BioTISR, independently of the biological content of each stack. Paired high-resolution references, where available, are excluded from the training objective.

\begin{table}[htbp]
\centering
\caption{Composition of the pretraining corpus.}
\label{tab:corpus}
\begin{tabular}{lrr}
\hline
Source & Training & Test \\
\hline
BioSR~\cite{qiao2021evaluation}   & \(140\)      & \(88\) \\
Synthetic                          & \(20{,}000\) & \(200\) \\
BioTISR~\cite{qiao2026neural}      & \(3500\)     & \(2400\) \\
\hline
Total                              & \(23{,}640\) & \\
\hline
\end{tabular}
\end{table}

Table~\ref{tab:optics_data} lists the optical parameters of the pretraining and test data.

\begin{table*}[t]
\centering
\caption{Optical parameters of the pretraining and test data. NA\(\cdot\Delta x\) is the product of the numerical aperture and the detector pixel size \(\Delta x\); at a given wavelength, the PSF width in pixels scales inversely with it. Exc., excitation; em., emission.}
\label{tab:optics_data}
\footnotesize
\setlength{\tabcolsep}{4.5pt}
\begin{tabular}{llcccc}
\hline
Data & Use & NA & Wavelength (nm) & \(\Delta x\) (nm) & NA\(\cdot\Delta x\) (nm) \\
\hline
BioSR & Pretraining & 1.70 (1.30 detection)\(^{a}\) & 488 exc. & 62.6 & 81.4\(^{a}\) \\
BioTISR, level 03 & Pretraining & 1.49 & 488, 561 exc. & 61.2 & 91.2 \\
Synthetic & Pretraining & 1.30 & 488 em. & 62.6 & 81.4 \\
Synthetic held-out & Test & 1.30 & 640, 405 em. & 62.6 & 81.4 \\
H1 (simulated) & Test & 1.49 & 640, 405 em. & 65.0 & 96.9 \\
H2 (simulated) & Test & 1.10 & 640, 405 em. & 55.0 & 60.5 \\
H3 (simulated) & Test & 1.49\(^{b}\) & 640, 405 em. & 65.0 & 96.9 \\
Stand microtubules, mitochondria & Test & 1.49\(^{c}\) & 525 em.\(^{c}\) & 65.0\(^{c}\) & 96.9 \\
Stand Argolight & Test & 1.42\(^{c}\) & 527 em.\(^{c}\) & 78.6\(^{c}\) & 111.6 \\
PCA-SIM mitochondria A, B & Test & 1.40 & 488 exc., 525 em. & 65.0 & 91.0 \\
\hline
\end{tabular}
\par\smallskip{\footnotesize\raggedright \(^{a}\)Objective NA and effective detection NA from the imaging conditions released with BioSR; NA\(\cdot\Delta x\) uses the detection NA. \(^{b}\)With \(0.060\) waves RMS horizontal astigmatism and \(0.053\) waves RMS horizontal coma (\(0.08\) waves RMS combined). \(^{c}\)Nominal values of the FlexSIM dataset catalogue.\par}
\end{table*}

The synthetic subset samples a \(5\times4\) grid of defocus and blur settings. Defocus values are \(-1.0\), \(-0.5\), \(0.0\), \(0.5\), and \(1.0~\mu\mathrm{m}\); Gaussian blur parameters are \(1.2\), \(1.8\), \(2.4\), and \(3.0\) pixels. Each condition contains \(1000\) training stacks and \(10\) test stacks. Each stack consists of nine \(128\times128\) SIM frames simulated at an emission wavelength of \(488~\mathrm{nm}\), a numerical aperture of \(1.3\), a native pixel size of \(62.6~\mathrm{nm}\), and a mean photon count of \(10^3\). Positive and negative defocus of equal magnitude produce identical kernels under the simulator's symmetric Airy-plus-ring defocus approximation, so the grid realizes fewer distinct kernels than settings. Here the blur values specify the Gaussian smoothing FWHM (with \(\sigma=\mathrm{FWHM}/2.355\)) on the sixfold PSF-generation grid during kernel construction. Simulated emitters include curves, tubes, rings, polygons, spots, and mixed structures. Component-level ground truth is retained for evaluation and is excluded from the training objective.

All FlexSIM sources are excluded from pretraining, as are the PCA-SIM stacks of Section~\ref{sec:experimental_results}. The tokens of the synthetic held-out domains are absent from the pretraining corpus (Appendix~\ref{app:psfcond}). Pretraining therefore uses neither target raw stacks nor a target-specific conditioning entry. Of the \(23{,}640\) candidate pretraining stacks, \(55\) fail the calibration checks of Appendix~\ref{app:calibration}, leaving \(23{,}585\) stacks for training, \(3585\) of them experimental. Group-level sampling balances the larger, more uniform synthetic subset with the real data. Each epoch samples raw SIM stacks at a \(1{:}2\) real-to-synthetic ratio. The synthetic stacks are drawn evenly from the \(20\) synthetic datasets, and BioSR and BioTISR contribute equally to the real-data portion. Sampling uses replacement when necessary.

\subsection{Pretraining protocol}\label{app:protocol}
The reconstruction model uses the physics-informed Transformer architecture of our previous work~\cite{wang2025self}, pretrained as a masked autoencoder~\cite{he2022masked,feichtenhofer2022masked}. A twelve-block encoder and an eight-block decoder process spatiotemporal patches, followed by convolutional output heads. A softplus activation enforces a non-negative emitter output on the threefold reconstruction grid. The PSF branch maps its condition token to a \(49\times49\) kernel. The original light-pattern and smooth-background heads remain in the checkpoint but are bypassed in the forward model, which uses the calibrated light pattern and background-corrected measurements. The checkpoint contains approximately \(2.4\times10^{8}\) parameters. Both pretraining and adaptation optimize the emitter network and PSF branch.

Pretraining is initialized from the checkpoint of our previous network, trained by self-supervision solely on the \(140\)-stack BioSR training split. Random \(80\times80\) raw-image crops are reconstructed at threefold spatial sampling. Training comprises ten epochs of sampled examples, data-parallel over eight NVIDIA A800-SXM4-40GB GPUs, and took \(4.9\) hours.

The model uses a spatiotemporal patch size of \(3\times16\times16\), a low-rank channel dimension of \(32\), and a masked-autoencoder ratio of \(0.75\). Noise augmentation is applied during pretraining only: zero-mean Gaussian noise is added independently to every pixel of a stack, with a standard deviation equal to \(\sigma\) times the mean intensity of that stack, where \(\sigma\) is drawn once per stack from a uniform distribution on \([0,1]\). The noise is treated as a constant with respect to the gradient.

Optimization uses AdamW with global-gradient-norm clipping at \(1.0\), a constant learning rate of \(4\times10^{-5}\), and zero weight decay for the shared emitter network and PSF branch, together with a separate Adam optimizer at learning rate \(10^{-2}\) for the per-stack light-pattern residual bank; the condition tokens remain fixed. Training progress is monitored on a fixed subset of training stacks; held-out synthetic domains and external experimental stacks are reserved strictly for evaluation.

Evaluation uses PSF condition tokens that are not updated during adaptation and carry no target-specific information (Appendix~\ref{app:psfcond}). For each scene, zero-shot and one-shot reconstruction receive the same token. For zero-shot reconstruction, the emitter is predicted directly from the raw stack in a single forward pass. The PSF branch contributes only to the optimization objective, which is not evaluated in this pass, so the token cannot affect the zero-shot emitter output.

\section{Adaptation objective and settings}
\subsection{Adaptation objective}\label{app:adapt-obj}
In single-stack adaptation, the same unlabeled target stack is used for self-supervised optimization and subsequent reconstruction; cross-stack reuse of the adapted weights on independent acquisitions is evaluated separately in Section~\ref{sec:experimental_results} and Appendix~\ref{app:additional}. For synthetic data, ground truth is used only to score the reconstruction after adaptation and is excluded from parameter updates.

For a target stack \(\mathbf{x}_j\), preprocessing applies the acquisition-specific background correction followed by full-stack percentile normalization, \(\mathbf{y}_j=\mathcal{N}[\mathcal{B}(\mathbf{x}_j)]\), which subtracts the \(0.01\mathrm{st}\) percentile and divides by the \(99\mathrm{th}\) percentile of the shifted stack. The light-pattern anchor is estimated from the same stack, normalized per frame to unit mean, and rendered on the threefold grid. The bounded residual \(\boldsymbol{\delta}_j\) allows adjustments to this anchor. For adapted network parameters \(\Phi\), the forward model in \eqref{eq:sim_forward_frame} is evaluated by fast Fourier transform:
\begin{equation}
    \tilde{\mathbf{y}}_{j,t}
    =
    \mathcal{D}_{\mathrm{area}}
    \left\{
        \mathcal{F}^{-1}
        \left[
            \mathcal{F}
            \left(
                \mathbf{h}_{j,\Phi}
            \right)
            \odot
            \mathcal{F}
            \left(
                \mathbf{e}_{j,\Phi}
                \odot
                T\,\widehat{\mathbf{P}}_{j,t}
            \right)
        \right]
    \right\},
    \label{eq:target_forward_model}
\end{equation}
where \(\mathbf{e}_{j,\Phi}\) and \(\mathbf{h}_{j,\Phi}\) are the target emitter and PSF predicted with the parameters \(\Phi\), and \(\mathcal{F}\) denotes the Fourier transform. The network light-pattern output is bypassed and no background term is added. The complete target objective is
\begin{equation}
\begin{aligned}
    \mathcal{L}_{\mathrm{tar}}^{(j)}
    &=
    \mathcal{L}_{\mathrm{rec}}^{(j)}
    +
    \lambda_h
    \mathcal{R}_{\mathrm{grad}}
    \left(\mathbf{h}_{j,\Phi}\right)
    +
    \lambda_P
    \mathcal{R}_{\mathrm{grad}}
    \left(\widehat{\mathbf{P}}_{j}\right)
    \\
    &\quad+
    \lambda_c
    \mathcal{R}_{\mathrm{center}}
    \left(\mathbf{h}_{j,\Phi}\right)
    +
    \lambda_{\delta}
    \mathcal{R}_{\delta}
    \left(\boldsymbol{\delta}_j\right)
    +
    \lambda_d
    \mathcal{R}_{\mathrm{dark}}
    \left(\mathbf{e}_{j,\Phi};\mathbf{y}_j\right).
\end{aligned}
    \label{eq:target_objective}
\end{equation}
The terms preceding the dark-region penalty have the same definitions as in Appendix~\ref{app:objective-full}, with masked reconstruction disabled and \(\mathbf{M}_j\) set to one. Target adaptation adds the dark-region term with weight \(\lambda_d=3\times10^{-3}\) and raises the PSF-gradient weight to \(\lambda_h=2\times10^{-3}\). No supervised image-domain loss is used.

The dark-region emitter penalty is
\begin{equation}
\begin{aligned}
    \mathcal{R}_{\mathrm{dark}}
    \left(\mathbf{e};\mathbf{y}\right)
    &=
    \frac{
        \sum_{\mathbf{u}}
        \omega_j(\mathbf{u})
        \left|\mathbf{e}(\mathbf{u})\right|
    }{
        \max\left\{
            \sum_{\mathbf{u}}\omega_j(\mathbf{u}),1
        \right\}
    },
    \\
    \omega_j(\mathbf{u})
    &=
    \operatorname{stopgrad}
    \left\{
        \exp\left[
            -\bar{\mathbf{y}}_j(\mathbf{u})/\tau
        \right]
    \right\}.
\end{aligned}
\label{eq:dark_penalty}
\end{equation}
Here \(\mathbf{u}\) indexes pixels of the emitter grid, \(\bar{\mathbf{y}}_j =\mathcal{U}_3[T^{-1}\sum_t\max(\mathbf{y}_{j,t},0)]\) is the frame-averaged measurement, and \(\mathcal{U}_3\) denotes linear resizing to the emitter grid. The decay parameter \(\tau=0.1\) is defined on the percentile-normalized intensity scale: the weight decreases to \(1/e\) at a normalized intensity of \(0.1\). The operator \(\operatorname{stopgrad}\) detaches these weights during back-propagation. As the weights decrease continuously with measured intensity, the penalty favors suppression in darker regions. As verified in Section~\ref{sec:experimental_results} and Appendix~\ref{app:adapt-settings}, setting \(\lambda_d=0\) leaves both synthetic accuracy and experimental line-pair resolution essentially unchanged, confirming that faint emitter structures are preserved.

\subsection{Adaptation settings and target scenarios}\label{app:adapt-settings}
Every primary adaptation comparison uses the same fixed budget of \(1200\) updates, for each synthetic test stack and for each experimental scene. On one NVIDIA A800-SXM4-40GB device, a run took a median of \(5.9\) minutes over the \(23\) stacks and at most \(13.8\) minutes (Stand mitochondria), including model loading, compilation, and the zero-shot and final reconstructions.

Each run samples random \(80\times80\) raw crops from the target stack with a mask ratio of zero. In full-network adaptation, the emitter network and PSF branch are fine-tuned with AdamW at learning rate \(10^{-5}\), global gradient-norm clipping at \(1.0\), and batch size \(1\), while the bounded 24-parameter light-pattern residual is updated by a separate Adam optimizer at learning rate \(10^{-2}\) with gain-logit, phase, and wave-vector limits of \(0.05\), \(0.15\) radians, and \(0.01\). The reconstruction is formed on the threefold-upsampled grid and mapped to the detector grid by finite-pixel integration.

\begin{figure*}[t]
\centering
\includegraphics[width=\textwidth]{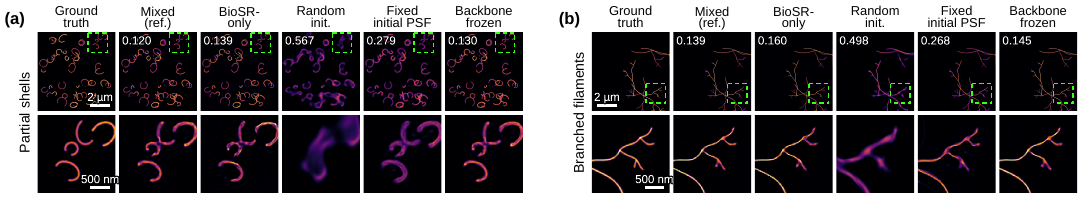}
\caption{Synthetic reconstructions of the controlled comparisons of Table~\ref{tab:ablation} after \(1200\) updates (columns: ground truth, mixed-condition reference, BioSR-only pretraining, random initialization, PSF fixed at its initial prediction, and frozen backbone) for representative (a) partial-shell (\(640~\mathrm{nm}\)) and (b) branched-filament (\(405~\mathrm{nm}\)) stacks, matching Fig.~\ref{fig:synthetic_transfer}. Top row of each domain: full \(8.0~\mu\mathrm{m}\) field with per-stack emitter NRMSE and enlarged region marked (dashed box); bottom row: enlarged \(2.0~\mu\mathrm{m}\) region. All reconstructions are intensity-aligned to the normalized ground truth on a shared \([0,1]\) scale. Scale bars, \(2~\mu\mathrm{m}\) (full field) and \(500~\mathrm{nm}\) (enlarged).}
\label{fig:s_ablation_syn}
\end{figure*}

\begin{figure}[t]
\centering
\includegraphics[width=\columnwidth]{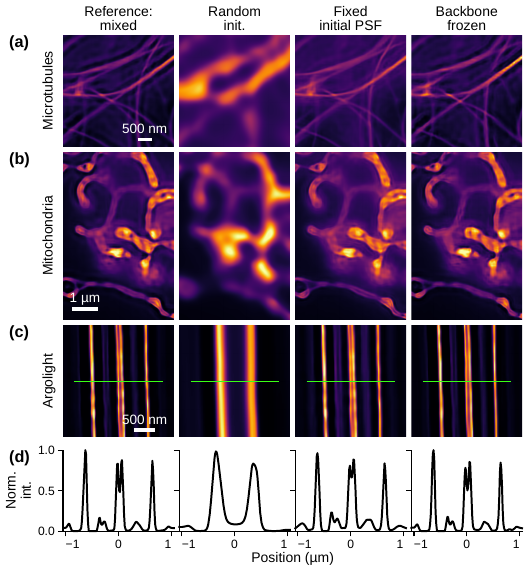}
\caption{Experimental reconstructions of the controlled comparisons of Table~\ref{tab:ablation} after \(1200\) updates (columns: mixed-condition reference, random initialization, PSF fixed at its initial prediction, and frozen backbone) in matched regions of (a) the microtubule, (b) the mitochondria, and (c) the Argolight scenes. Each image is displayed between its full-field 0.1st and 99.9th percentiles. (d) Profiles along the lines in (c) across the nominal \(60~\mathrm{nm}\) group marked in Fig.~\ref{fig:external_transfer}c (central double line between two single reference lines), sampled on the \(3072\times3072\) display grid after normalizing by its full-field 0.1st and 99.9th percentiles. Random initialization gives no central peak; the mixed-condition reference, fixed-PSF, and frozen-backbone variants each show one peak at each single line and two peaks at the central double line (minimum-to-peak ratios \(0.565\), \(0.782\), and \(0.614\)).}
\label{fig:s_ablation_real}
\end{figure}

\begin{figure*}[t]
\centering
\includegraphics[width=\textwidth]{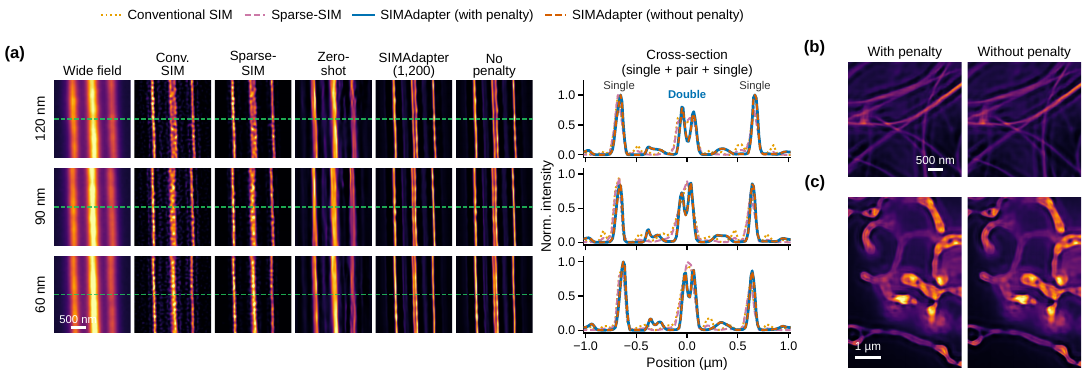}
\caption{Argolight calibration-target verification and dark-region penalty ablation at \(1200\) updates. (a) Reconstructed regions of the Stand Argolight target at nominal \(120\), \(90\), and \(60~\mathrm{nm}\) line-pair spacings (rows, \(100\times100\) samples on the threefold grid, \(2.62\times2.62~\mu\mathrm{m}\), centered at \(y=600\); scale bar \(500~\mathrm{nm}\)), each containing one flanking single reference line on each side of a central double line pair across six reconstruction columns (wide field, conventional SIM, Sparse-SIM, SIMAdapter zero-shot, SIMAdapter at \(1200\) updates with \(\lambda_d=0.003\), and SIMAdapter without the dark-region penalty, \(\lambda_d=0\)), paired with cross-section profiles along the dashed green line for the four methods listed in the legend. SIMAdapter distinguishes flanking single reference lines from central double line pairs across the tested nominal \(120\), \(90\), and \(60~\mathrm{nm}\) spacings, with and without the penalty. (b) Microtubule and (c) mitochondria regions with (left) and without (right) the penalty; each pair shares one display range computed from both full fields.}
\label{fig:s_dark}
\end{figure*}

The variants of Table~\ref{tab:ablation} share these settings and differ in one factor each. The BioSR-only variant starts from the checkpoint of our previous network that initializes pretraining (Appendix~\ref{app:protocol}). The randomly initialized variant uses default random weight initialization for the architecture and the same fixed tokens. The fixed-PSF variant holds the initial kernel predicted before adaptation fixed in the forward model and freezes the PSF branch. The backbone-frozen variant freezes the shared masked-autoencoder encoder and decoder (\(1.88\times10^{8}\) parameters) and updates the emitter readout, low-rank, and physics branches (\(5.23\times10^{7}\) parameters) with their learning rates unchanged. The variant without the dark-region penalty sets \(\lambda_d=0\). Figures~\ref{fig:s_ablation_syn}, \ref{fig:s_ablation_real}, and~\ref{fig:s_dark} show the synthetic and experimental reconstructions of these variants.

\begin{figure}[t]
\centering
\includegraphics[width=\columnwidth]{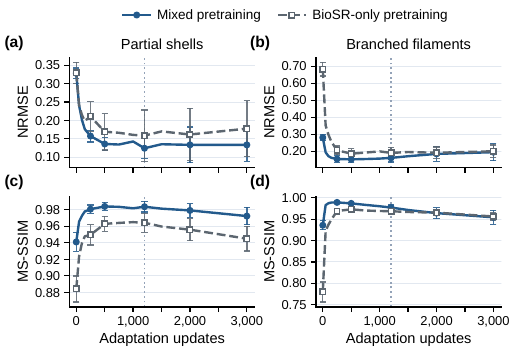}
\caption{Synthetic adaptation curves for the mixed-condition and BioSR-only networks: emitter (a, b) NRMSE and (c, d) MS-SSIM of partial shells (a, c) and branched filaments (b, d) after \(0\) to \(3000\) updates. Lines show means over ten stacks per domain, error bars show one standard deviation at selected checkpoints (\(0\), \(250\), \(500\), \(1200\), \(2000\), and \(3000\) updates), and the dotted line marks the budget of \(1200\) updates. The curves come from separate \(3000\)-update runs; Fig.~\ref{fig:synthetic_transfer} uses separate \(1200\)-update runs.}
\label{fig:s_iterations}
\end{figure}

\begin{figure}[t]
\centering
\includegraphics[width=\columnwidth]{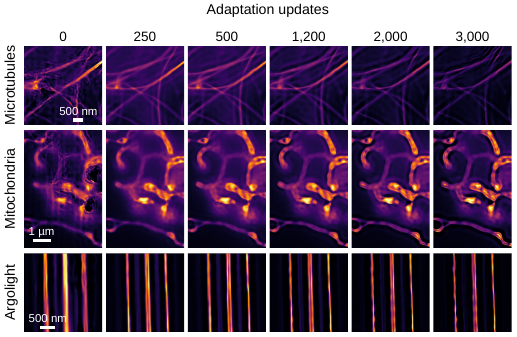}
\caption{Fixed regions of the microtubule, mitochondria, and Argolight scenes (rows, matched to Fig.~\ref{fig:external_transfer}) after \(0\) to \(3000\) adaptation updates (columns). Each row comes from a single extended run, separate from the \(1200\)-update runs of Fig.~\ref{fig:external_transfer}. Each image is displayed between its full-field 0.1st and 99.9th percentiles.}
\label{fig:s_steps}
\end{figure}

Figure~\ref{fig:s_iterations} shows the synthetic accuracy after \(0\) to \(3000\) updates, and Fig.~\ref{fig:s_steps} the experimental reconstructions after \(0\) to \(3000\) updates, from runs separate from the \(1200\)-update reconstructions of the main text. Over the \(20\) synthetic stacks, the mean NRMSE of the mixed-condition network was \(0.304\), \(0.156\), \(0.144\), \(0.143\), and \(0.164\) after \(0\), \(250\), \(500\), \(1200\), and \(3000\) updates.
The synthetic targets are partial shells at \(640~\mathrm{nm}\) and
branched filaments at \(405~\mathrm{nm}\). Each domain contains ten
independently generated stacks of nine \(128\times128\) frames. Emission
wavelength and emitter morphology differ from synthetic pretraining;
numerical aperture, pixel size, defocus, and blur settings are unchanged.
The simulator uses a numerical aperture of \(1.3\), a native pixel size
of \(62.6~\mathrm{nm}\), emission wavelengths of \(640\) and
\(405~\mathrm{nm}\), an illumination period of five detector pixels, and
a modulation depth of \(0.8\). The PSF has zero defocus and a Gaussian FWHM blur parameter of \(1.8\) pixels on the sixfold generation grid, both selected from the pretraining grid. Stacks
are scaled to \(10^{3}\) foreground photons and given Poisson shot noise, a
signal-dependent Gaussian term at \(0.1\) times the photon-image standard
deviation, a \(500\)-count offset, and \(8\) counts of read noise, then
rescaled to \([0,1]\). The resulting background retains a measurable
intensity distribution without truncation at zero.

The experimental scenes are Stand microtubules, Stand mitochondria, and Stand Argolight from FlexSIM. Each provides one nine-frame acquisition without a paired emitter reference and is evaluated individually.

\subsection{Additional tests}\label{app:additional}
\begin{figure*}[t]
\centering
\includegraphics[width=\textwidth]{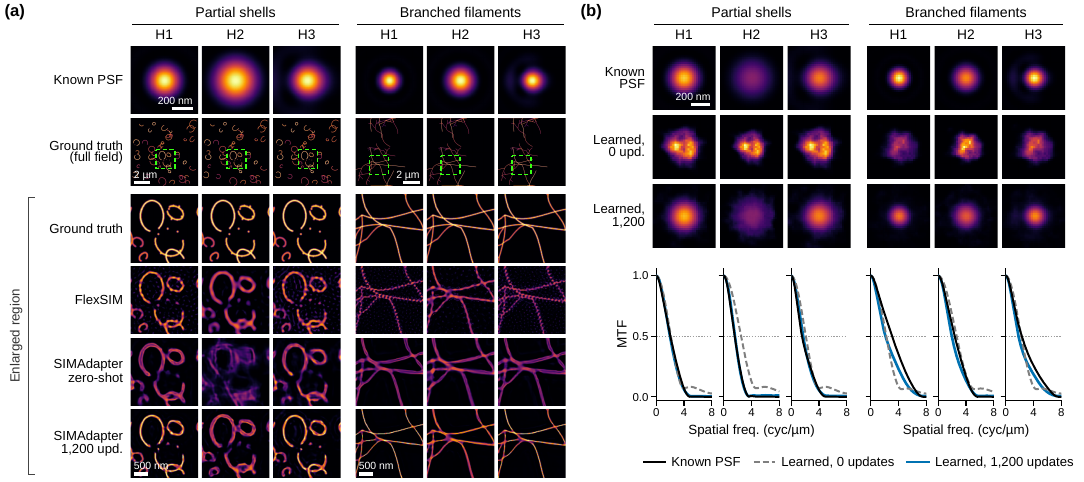}
\caption{Simulated configurations H1--H3 (Table~\ref{tab:optics}) for a representative test stack of each domain (partial shells, left three columns; branched filaments, right three columns). (a) Rows, top to bottom: known PSF (scale bar \(200~\mathrm{nm}\)), full-field ground truth with the enlarged region marked by a dashed box (scale bar \(2~\mu\mathrm{m}\)), and enlarged views of ground truth, FlexSIM, SIMAdapter zero-shot, and SIMAdapter after \(1200\) updates (scale bar \(500~\mathrm{nm}\)); emitter panels are aligned in intensity to the ground truth and share one display range. (b) Known PSF integrated onto the grid of the learned kernel, learned kernels after \(0\) and \(1200\) updates (one display maximum per column), and radial MTFs, in the same columns as (a).}
\label{fig:s_hardware}
\end{figure*}

\begin{figure*}[t]
\centering
\includegraphics[width=\textwidth]{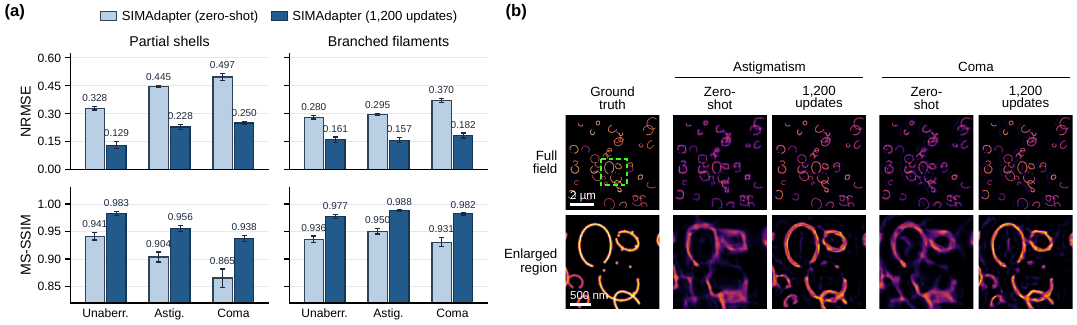}
\caption{Pupil aberrations of \(0.10\) waves RMS. (a) Mean NRMSE (top) and MS-SSIM (bottom) of the ten partial-shell (left) and ten branched-filament (right) stacks before (light blue) and after \(1200\) updates (dark blue) for the unaberrated pupil, horizontal astigmatism, and horizontal coma, with \(95\%\) bootstrap intervals; the unaberrated partial-shell value (\(0.129\)) is from the paired clear-pupil control run, whereas Table~\ref{tab:optics} lists the main \(1200\)-update run (\(0.128\)). Partial shells under coma are scored after independently aligning each reconstruction to the ground truth by subpixel phase correlation with upsampling factor \(50\) (\(0.497\rightarrow0.250\) NRMSE, \(0.865\rightarrow0.938\) MS-SSIM; pixel-locked NRMSE is \(0.433\rightarrow0.433\) owing to the \(28.8\pm2.5~\mathrm{nm}\) lateral translation along the coma axis). (b) Representative partial-shell stack under horizontal astigmatism and horizontal coma: full-field views (top row; scale bar \(2~\mu\mathrm{m}\)) and enlarged views of the dashed region (bottom row; scale bar \(500~\mathrm{nm}\)) for ground truth, zero-shot, and \(1200\)-update adapted reconstructions (displayed with the mean \(-1.38\)-pixel shift removed under coma).}
\label{fig:s_aberration}
\end{figure*}

The tests of unseen optical conditions (Section~\ref{sec:optics_results}; Figs.~\ref{fig:s_hardware} and \ref{fig:s_aberration}) reuse the \(20\) held-out emitters and the detector model above. Aberrated PSFs are computed from a circular pupil with horizontal Zernike astigmatism or coma of \(0.10\) waves RMS (and \(0.20\) waves RMS to probe the stronger-aberration boundary). H1--H3 change the numerical aperture, the pixel size, and, for H2, the illumination period; H3 adds \(0.060\) waves RMS horizontal astigmatism and \(0.053\) waves RMS horizontal coma (\(0.08\) waves RMS combined, Strehl ratio \(0.78\)). Their kernels are computed at six times the raw sampling on a \(129\times129\) support, which retains \(91\)--\(97\%\) of the kernel energy before normalization. The expected nine-frame photon count of each stack equals that of its reference stack, and noise uses the same seeds. Every stack is calibrated from its raw frames with FlexSIM. For H1--H3, all methods use FlexSIM carrier estimation with the wide-field mask set to \(0\). With the default mask of \(0.2\), the estimated light pattern correlated with the simulated illumination below \(0.90\) on \(13\) of the \(60\) stacks (filaments; \(6\) in H1 and \(7\) in H3); with the mask set to \(0\), the stack-mean correlation was at least \(0.983\) on all \(60\) stacks. The clear-pupil reference at the pretraining NA and pixel size uses the default mask.

\begin{figure*}[t]
\centering
\includegraphics[width=\textwidth]{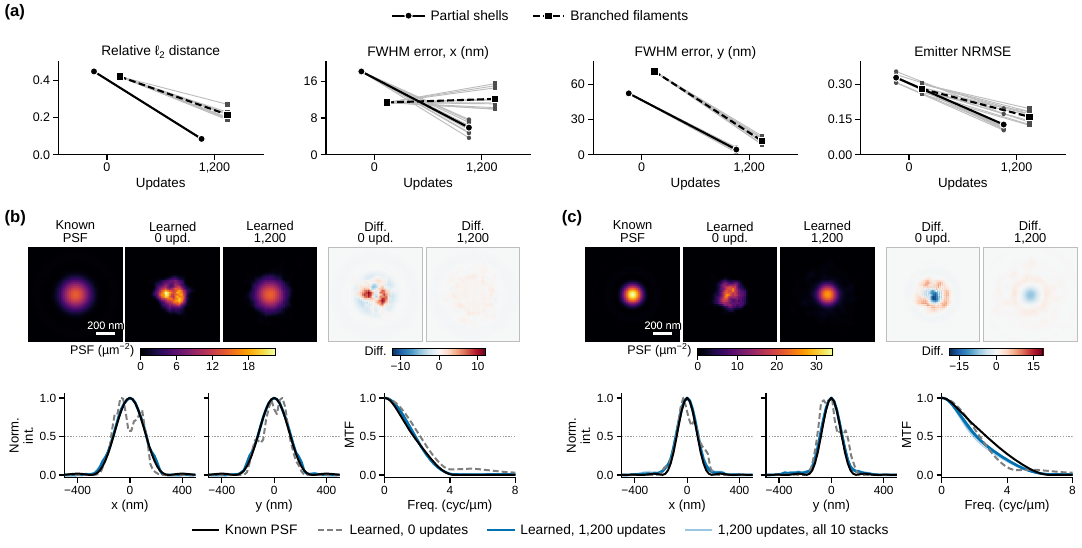}
\caption{Learned PSFs compared with the known generating kernels. (a) Per-stack values at \(0\) and \(1200\) updates of the centered relative \(\ell_2\) distance, the absolute FWHM errors in x and y of the learned kernel, and the emitter NRMSE. Gray markers and lines show individual stacks (circles, partial shells; squares, branched filaments), and black markers the domain means; at \(0\) updates the ten kernels of a domain are identical. (b, c) First stack of the partial-shell (b) and branched-filament (c) domains on the common \(20.9~\mathrm{nm}\) grid. Top row: known PSF, learned PSFs after \(0\) and \(1200\) updates, and learned minus known after centroid alignment; scale bar, 200~nm. Bottom row: x and y profiles through the centroid and radial MTF; thin light lines show all ten stacks after \(1200\) updates.}
\label{fig:s_psf}
\end{figure*}

\begin{figure}[t]
\centering
\includegraphics[width=\columnwidth]{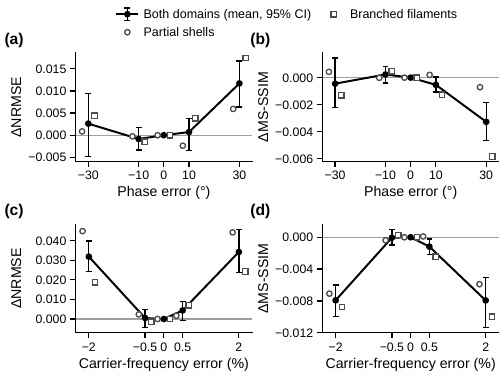}
\caption{Sensitivity to calibration errors. Change of (a, c) NRMSE and (b, d) MS-SSIM after \(1200\) updates when the phase (a, b) or the carrier frequency (c, d) of one orientation deviates from the calibrated value, relative to the same stack reconstructed with the unperturbed calibration. Filled markers show the mean over the 20 synthetic stacks with \(95\%\) bootstrap intervals; open markers show the means of the ten partial-shell and the ten branched-filament stacks, offset horizontally for clarity.}
\label{fig:s_calibration}
\end{figure}

\begin{figure*}[t]
\centering
\includegraphics[scale=0.92]{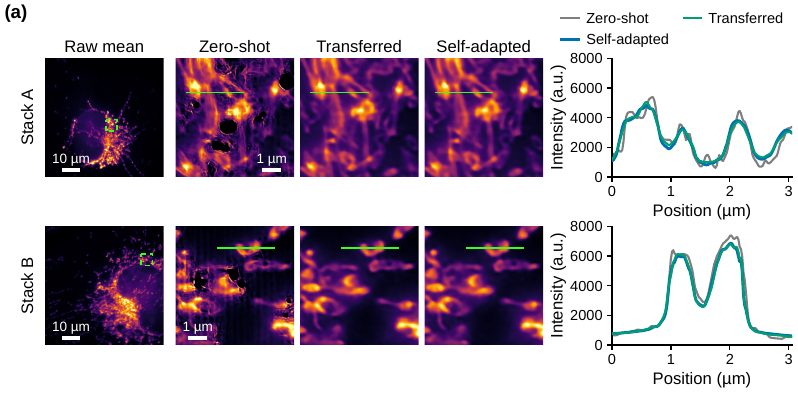}\\[4pt]
\includegraphics[scale=0.92]{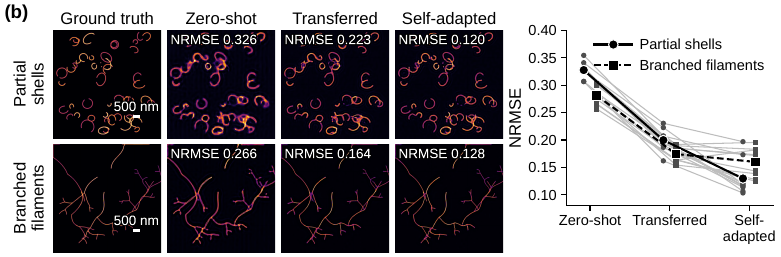}
\caption{Cross-stack reuse of adapted SIMAdapter weights. (a) Two separately recorded mitochondria stacks (stacks A, top row, and B, bottom row) from the PCA-SIM microscope~\cite{qian2023structured}: mean raw frame with the enlarged region marked; the region reconstructed zero-shot, by SIMAdapter adapted to the other stack and applied without updates (transferred), and after \(1200\) updates on the stack itself (self-adapted); right, profiles along the marked lines. The three reconstructions of each stack share one display range. (b) Synthetic domains (top row, partial shells; bottom row, branched filaments): ground truth and the zero-shot, transferred, and self-adapted reconstructions of the second stack of each domain, with the NRMSE of each reconstruction; the transferred network is the one adapted to the first stack. Right, NRMSE of the other nine stacks of both domains for the three cases, with gray lines joining each stack and black markers for the domain means.}
\label{fig:s_cross}
\end{figure*}

Figures~\ref{fig:s_psf}, \ref{fig:s_calibration}, and~\ref{fig:s_cross} show the PSF comparison, the calibration-error tests, and the cross-stack test. The calibration-error tests (Fig.~\ref{fig:s_calibration}) shift the phase at the detector center or the magnitude of the carrier wave vector for one orientation of each stack, assigned in rotation across stacks, in both the light-pattern image and its parameters; the bounded residual remains active. For the PSF comparison (Fig.~\ref{fig:s_psf}), each true kernel (\(99\times99\) samples at the nominal \(10.4~\mathrm{nm}\) pitch) is integrated onto the physical support of the predicted kernel (\(49\times49\) samples at \(20.9~\mathrm{nm}\)), both are normalized to unit sum and centered at their centroids, and the centered relative \(\ell_2\) distance is the norm of their difference divided by the norm of the true kernel. The cross-stack test (Fig.~\ref{fig:s_cross}) adapts each PCA-SIM stack for \(1200\) updates, transfers the network weights and fixed tokens to the other stack, and evaluates it without updates using the calibrated light pattern of the target. The relative residual between reprojected and measured frames is computed for all three networks of a target with its calibrated light pattern, a common PSF, zero background, and one fitted gain per reconstruction. On the synthetic domains, the network adapted to the first test stack of each domain for \(1200\) updates is applied without updates to the remaining nine stacks, each with its own calibrated light pattern and a zero residual, and the emitters are scored against their references.

\section{Baselines, metrics, and image analysis}

\subsection{Baseline reconstruction parameters}\label{app:baselines-detail}
Wide-field images are computed by averaging the nine raw frames.

Conventional SIM differs between the two benchmarks. On synthetic data it uses the simulator's frequency-domain Wiener reconstruction with the known carrier orientation, light-pattern period, modulation depth, and transfer function; only the starting phase of each of the three orientations is estimated from the raw frames, providing a simulator-informed reference. On experimental data it uses blind FlexSIM reconstruction, which estimates the light pattern from each measured stack and produces a native twofold output without simulation parameters.

Sparse-SIM is applied to the conventional SIM output in both benchmarks and inherits whatever information that reconstruction used. Its parameters are selected by one procedure throughout, following the visual adjustment prescribed by the method's user manual: candidate settings spanning the fidelity, sparsity, background-estimation and deconvolution axes are reconstructed on calibration stacks and inspected, and the setting retained is the one that sharpens structures without fragmenting continuous features into disconnected pieces. The calibration stacks are disjoint from the evaluated ones and match them in morphology: for the experimental benchmark they are three FlexSIM scenes that are not evaluated here, and for the synthetic benchmark they are stacks regenerated from the same simulator configuration under a different random seed. No emitter ground truth enters this selection. One parameter set is retained for the synthetic benchmark and one for the experimental benchmark. Each set is held fixed across all evaluated stacks within its benchmark.

\begin{figure*}[t]
\centering
\includegraphics[scale=0.94]{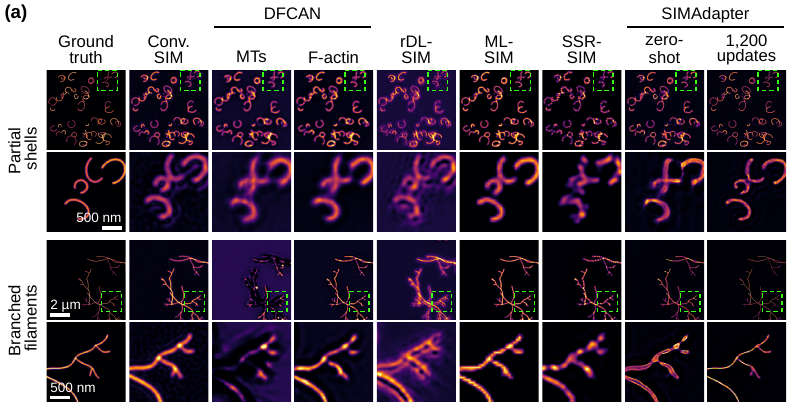}\\[4pt]
\includegraphics[width=\textwidth]{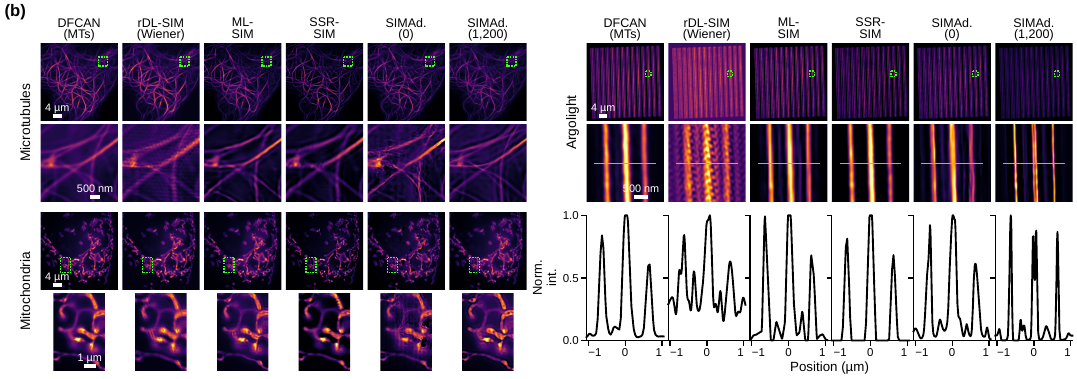}
\caption{External learned reconstructions. (a) One partial-shell stack (top pair of rows) and one branched-filament stack (bottom pair of rows), each shown as the full \(8.0~\mu\mathrm{m}\) field with the enlarged region marked (dashed box; scale bar \(2~\mu\mathrm{m}\)) and as the enlarged \(2.0~\mu\mathrm{m}\) region (scale bar \(500~\mathrm{nm}\)): ground truth and reconstructions by conventional SIM, DFCAN trained on BioSR microtubules (MTs) or F-actin, rDL-SIM with the Wiener back end, ML-SIM, SSR-SIM, and SIMAdapter zero-shot and after \(1200\) adaptation updates. (b) External experimental scenes reconstructed by DFCAN trained on microtubules, rDL-SIM with the Wiener back end, ML-SIM, SSR-SIM, and SIMAdapter zero-shot and after \(1200\) updates (left: microtubules and mitochondria; right: Argolight, with profiles plotted below the images along the line marked in Fig.~\ref{fig:external_transfer}c, normalized to each profile's maximum). Each scene shows the full field with the enlarged region marked and the enlarged region below, scaled between its 0.1st and 99.9th percentiles. External learned baselines use released weights without target updates; SIMAdapter is shown before and after \(1200\) updates.}
\label{fig:s_external}
\end{figure*}

On synthetic data, FlexSIM estimates the light pattern from the raw frames and uses an analytical transfer function at the nominal NA, wavelength, and pixel size, producing a threefold output. The learned references are evaluated without target updates: released DFCAN models trained on BioSR microtubules and F-actin with the published post-processing~\cite{qiao2021evaluation} (plus an author-trained BioSR CCPs model in Table~\ref{tab:all_baselines}); rDL-SIM, whose released microtubule networks denoise the nine raw frames before FlexSIM or Wiener reconstruction using the raw-calibrated light pattern and nominal optical metadata~\cite{qiao2023rationalized}; ML-SIM, trained on randomized synthetic SIM images~\cite{christensen2021ml}; and SSR-SIM, trained by self-supervision on LifeAct data~\cite{liu2025bio}, which receives the transfer function of the simulator PSF on synthetic data and a rescaled measured transfer function on experimental data. DFCAN and SSR-SIM produce twofold outputs, ML-SIM a onefold output, and rDL-SIM a threefold output. Table~\ref{tab:synthetic} lists rDL-SIM with the FlexSIM back end, which was more accurate than the Wiener back end (Table~\ref{tab:all_baselines}). Figure~\ref{fig:s_external} shows the reconstructions of all learned methods on the synthetic stacks and the experimental scenes, with rDL-SIM shown with the Wiener back end.

\begin{table}[t]
\centering
\caption{All reference reconstructions on the two synthetic domains. NRMSE / MS-SSIM are means over ten stacks per domain; experimental reconstructions and Argolight line-pair profiles are shown in Fig.~\ref{fig:s_external}b.}
\label{tab:all_baselines}
\footnotesize
\setlength{\tabcolsep}{3.5pt}
\begin{tabular}{lcc}
\hline
Method & Partial shells & Branched filaments \\
\hline
Conventional SIM\(^{a}\) & 0.442 / 0.902 & 0.426 / 0.922 \\
Wiener-SIM, raw calibration & 0.590 / 0.809 & 0.541 / 0.833 \\
FlexSIM, raw calibration\(^{b}\) & 0.253 / 0.952 & 0.412 / 0.928 \\
Sparse-SIM & 0.414 / 0.915 & 0.391 / 0.933 \\
DFCAN (BioSR CCPs)\(^{c}\) & 0.578 / 0.860 & 0.631 / 0.854 \\
DFCAN (microtubules) & 0.509 / 0.887 & 0.896 / 0.663 \\
DFCAN (F-actin) & 0.412 / 0.919 & 0.365 / 0.942 \\
rDL-SIM, Wiener back end & 0.612 / 0.817 & 0.606 / 0.859 \\
rDL-SIM, FlexSIM back end & 0.336 / 0.940 & 0.563 / 0.886 \\
ML-SIM & 0.316 / 0.946 & 0.314 / 0.955 \\
SSR-SIM & 0.506 / 0.882 & 0.463 / 0.917 \\
SIMAdapter, zero shot & 0.328 / 0.941 & 0.280 / 0.936 \\
SIMAdapter, 1200 updates & 0.128 / 0.983 & 0.161 / 0.977 \\
\hline
\end{tabular}
\par\smallskip{\scriptsize\raggedright \(^{a}\)Wiener reconstruction with the simulated light-pattern parameters. \(^{b}\)Threefold FlexSIM reconstruction from the raw frames. \(^{c}\)Trained by the authors on BioSR clathrin-coated pits.\par}
\end{table}

The zero-shot reference uses the mixed-pretraining checkpoint that initializes every primary adaptation run. It predicts the emitter in one forward pass without a parameter update, on the same threefold reconstruction grid used for adaptation. The emitter prediction uses neither the physical forward model nor light-pattern calibration. Both reconstructions use the same PSF token. Because the PSF branch contributes only to the optimization objective, its token cannot affect the zero-shot emitter prediction. All methods use matched physical fields of view. Outputs are resampled as needed according to their physical pixel calibration. Display normalization follows quantitative evaluation.

\subsection{Metric definitions and implementation}\label{app:metrics-detail}
All synthetic reconstructions, including wide-field and conventional SIM, are resampled to \(384\times384\) pixels over the same physical field of view, at three times the \(128\times128\) raw sampling. Two pipelines evaluate NRMSE and MS-SSIM. For NRMSE, both images are first convolved with a Gaussian kernel of \(\sigma=1.5\) output pixels under reflecting boundary handling, then normalized independently by mapping their 1st and 99.8th percentiles to zero and one with clipping outside that range; a least-squares scale and offset align the normalized reconstruction to the normalized ground truth without clipping the fitted residual, and NRMSE is the \(\ell_2\) norm of their difference divided by the \(\ell_2\) norm of the smoothed ground truth. For MS-SSIM, the same independent 1st/99.8th percentile normalization and least-squares affine alignment are applied without pre-smoothing, with the aligned image clipped to \([0,1]\). Both metrics are averaged over the ten test stacks of each synthetic domain. MS-SSIM uses the five-scale measure of Wang \textit{et al.}~\cite{wang2003multiscale}, with standard scale weights and an \(11\times11\) Gaussian window of \(\sigma=1.5\). Local statistics are computed with reflecting boundary handling and averaged over the full field. Averaging over a cropped valid region changes the pooled method means of the learned reconstructions and Sparse-SIM in Fig.~\ref{fig:synthetic_transfer}c,d by at most \(0.004\), while the wide-field mean decreases by \(0.013\). Each per-scale term is clamped at \(10^{-8}\) before exponentiation, as in common reference implementations; the clamp is inactive for all reported reconstructions (for the Fig.~\ref{fig:synthetic_transfer} methods, the smallest per-scale term is \(0.69\)).

Ground truth is excluded from pretraining and adaptation. All primary benchmark scores are computed from the final reconstruction at the fixed budget of \(1200\) updates; intermediate reconstructions are not selected by their evaluation scores. Bootstrap intervals resample stacks with replacement \(10{,}000\) times; when both domains are pooled, each resample keeps ten stacks per domain. Wilcoxon signed-rank tests are two-sided and exact, and the comparisons of random initialization, the fixed initial PSF, and the frozen backbone with the reference in Table~\ref{tab:ablation} are Holm-adjusted across these three tests.

\subsection{Decorrelation analysis}\label{app:decorr-detail}
We use the published estimator of Descloux \textit{et al.}~\cite{descloux2019parameter} with \(50\) radial sampling positions for each image. For a physical pixel size \(p\) and the normalized frequency \(k_{\mathrm{dec}}\) of the decorrelation peak, with \(k_{\mathrm{dec}}=1\) at the two-pixel sampling limit, the estimate is
\begin{equation}
    d_{\mathrm{dec}}
    =
    \frac{2p}{k_{\mathrm{dec}}}.
    \label{eq:decorrelation_resolution}
\end{equation}
Images are analyzed before display normalization, without sharpening or smoothing. Decorrelation estimates summarize apparent spatial-frequency support on each method's native output grid and are interpreted alongside direct line-pair verification on the Argolight calibration target (Section~\ref{sec:experimental_results}).

The native pixel sizes are \(65~\mathrm{nm}\) for Stand microtubules, \(65~\mathrm{nm}\) for Stand mitochondria, and \(78.6~\mathrm{nm}\) for Stand Argolight. The twofold outputs have pixel sizes of \(32.5\), \(32.5\), and \(39.3~\mathrm{nm}\); the corresponding values for the threefold SIMAdapter outputs are \(21.67\), \(21.67\), and \(26.2~\mathrm{nm}\). All methods are analyzed over matched physical fields of view.

Table~\ref{tab:decorrelation} lists the estimates.

\begin{table}[t]
\centering
\caption{Decorrelation length scales (nm) of the experimental reconstructions on their native output grids. The last column gives the range over five fixed subfields (four quadrants and the center, each half the field in height and width) of the adapted reconstruction; the number of subfields whose decorrelation peak lies above \(0.95\) of the Nyquist frequency is given in parentheses. The two-pixel sampling limit of the threefold grid is \(43.3~\mathrm{nm}\) for microtubules and mitochondria and \(52.4~\mathrm{nm}\) for Argolight.}
\label{tab:decorrelation}
\scriptsize
\setlength{\tabcolsep}{1.8pt}
\begin{tabular}{lcccccc}
\hline
Scene & WF & Conv. & Sparse & Zero & 1200 & Subfields \\
\hline
Microtubules & 254.8 & 95.4 & 109.8 & 52.2 & 47.8 & 44.8--49.8 (1) \\
Mitochondria & 277.0 & 105.4 & 116.8 & 51.1 & 54.3 & 44.3--52.2 (3) \\
Argolight & 232.8 & 85.2 & 78.8 & 62.2 & 60.1 & 52.6--62.1 (2) \\
\hline
\end{tabular}
\par\smallskip{\scriptsize\raggedright WF, wide field; Conv., conventional SIM; Sparse, Sparse-SIM; Zero, zero-shot; 1200, 1200 updates.\par}
\end{table}

\begin{table*}[t]
\centering
\caption{Full-board line-pair statistics on Stand Argolight across \(1200\) threefold-grid rows (\(y=200\text{--}1399\)) for the nominal \(120\), \(90\), and \(60~\mathrm{nm}\) single--double--single groups (Appendix~\ref{app:expcompare-detail}). Entries list the single-peak fraction on flanking reference lines (averaged over all three groups) and the two-peak fraction, median peak spacing (nm), and median minimum-to-mean-peak ratio for each central double line.}
\label{tab:linepairs}
\scriptsize
\setlength{\tabcolsep}{4.5pt}
\begin{tabular}{lcccc}
\hline
Method & Single lines (1-pk) & 120~nm (2-pk; sep; min/pk) & 90~nm (2-pk; sep; min/pk) & 60~nm (2-pk; sep; min/pk) \\
\hline
Wide field & 100.0\% & 0.0\%; -- & 0.0\%; -- & 0.0\%; -- \\
Conventional SIM\(^{a}\) & 50.6\% & 96.5\%; 157.2; 0.335 & 42.3\%; 117.9; 0.649 & 1.6\%\(^{b}\); 157.2; 0.180 \\
Sparse-SIM & 99.5\% & 93.6\%; 117.9; 0.375 & 26.2\%; 117.9; 0.686 & 0.0\%; -- \\
DFCAN (MTs) & 100.0\% & 0.0\%; -- & 0.0\%; -- & 0.0\%; -- \\
DFCAN (F-actin) & 100.0\% & 0.0\%; -- & 0.0\%; -- & 0.0\%; -- \\
DFCAN (CCPs) & 100.0\% & 0.0\%; -- & 0.0\%; -- & 0.0\%; -- \\
rDL-SIM (Wiener back end) & 78.3\% & 4.4\%; 131.0; 0.629 & 7.5\%; 117.9; 0.749 & 19.1\%\(^{b}\); 131.0; 0.718 \\
rDL-SIM (FlexSIM back end) & 100.0\% & 1.7\%; 183.4; 0.164 & 0.3\%; 176.8; 0.094 & 0.0\%; -- \\
ML-SIM & 99.7\% & 0.0\%; -- & 0.0\%; -- & 0.0\%; -- \\
SSR-SIM & 100.0\% & 0.0\%; -- & 0.0\%; -- & 0.0\%; -- \\
SIMAdapter, zero shot & 70.4\% & 1.6\%; 144.1; 0.687 & 3.3\%; 104.8; 0.609 & 5.1\%; 78.6; 0.827 \\
SIMAdapter, 1200 updates & \textbf{100.0\%} & \textbf{100.0\%; 131.0; 0.137} & \textbf{99.8\%; 104.8; 0.408} & \textbf{83.5\%; 78.6; 0.626} \\
\quad without dark-region penalty & 100.0\% & 100.0\%; 131.0; 0.149 & 97.8\%; 104.8; 0.418 & 82.2\%; 78.6; 0.634 \\
\quad random initialization & 0.0\% & 0.0\%; -- & 0.0\%; -- & 0.0\%; -- \\
\quad fixed initial PSF & 100.0\% & 100.0\%; 131.0; 0.303 & 93.2\%; 104.8; 0.597 & 44.6\%; 78.6; 0.757 \\
\quad backbone frozen & 100.0\% & 100.0\%; 117.9; 0.190 & 96.5\%; 91.7; 0.465 & 85.9\%; 78.6; 0.642 \\
\hline
\end{tabular}
\par\smallskip{\scriptsize\raggedright \(^{a}\)Twofold blind FlexSIM reconstruction of Fig.~\ref{fig:external_transfer}; flanking sidelobes split \(\sim\!49\%\) of single-line profiles. \(^{b}\)Periodic ringing/stripe peaks spaced \(131\)--\(157~\mathrm{nm}\) apart rather than the \(60~\mathrm{nm}\) pair.\par}
\end{table*}

\subsection{Experimental image comparison}\label{app:expcompare-detail}
Profiles are sampled along fixed line segments that are identical for all methods within each scene. Each image is first normalized by its full-field 0.1st and 99.9th percentiles. For Stand Argolight, the profile marked in Fig.~\ref{fig:external_transfer}c crosses the nominal \(60~\mathrm{nm}\) group along row \(y=600\) of the \(1536\times1536\) threefold grid, from \(x=1189\) to \(1269\), and spans both flanking single reference lines and the central double line; the enlarged \(100\times100\) region starts at \((x,y)=(1179,550)\) and matches the \(60~\mathrm{nm}\) row of Fig.~\ref{fig:s_dark}a. For the calibration-target verification across the nominal \(120\), \(90\), and \(60~\mathrm{nm}\) line-pair groups of the \(390\) to \(0~\mathrm{nm}\) Argo-SIM pattern, Fig.~\ref{fig:s_dark}a shows \(100\times100\) threefold-grid regions (\(2.62\times2.62~\mu\mathrm{m}\)) centered at \(y=600\) (\(x=1030\), \(1130\), and \(1229\)) and cross-sections spanning \(\pm 40\) threefold-grid samples (\(\pm 1.05~\mu\mathrm{m}\)) to include both flanking single reference lines and the central pair; full-board statistics scan all \(1200\) threefold-grid rows \(y=200\text{--}1399\) following the slight \(1.8^\circ\) board tilt, slicing each central pair within \(\pm 14\) display samples (\(\pm 183.4~\mathrm{nm}\)) of the rounded tilt-corrected group midpoint and each flanking single line within \(\pm 10\) threefold-grid samples (\(\pm 20\) display samples). Intensities are read on the \(3072\times3072\) display grid (spacing \(13.1~\mathrm{nm}\)) after full-field 0.1st--99.9th percentile normalization, and each sliced profile window is divided by its window maximum before peak detection. A peak qualifies if its height and prominence, both relative to the profile maximum, are at least \(0.15\) and \(0.10\), and if it lies at least four display samples from any other qualifying peak. When exactly two peaks qualify, we report their spacing and the ratio of the minimum between them to their mean height. Spacings are read at the display sampling. Table~\ref{tab:linepairs} lists the full-board statistics of all methods. The microtubule and mitochondria profiles describe local peak shape and are compared visually.

\section*{Back matter}
\begin{backmatter}
\bmsection{Funding}
Fang-Wen Sun and Xiang-Dong Chen were supported by National Natural Science Foundation of China (No. 62225506), CAS Project for Young Scientists in Basic Research (No. YSBR-049) and the Innovation Program for Quantum Science and Technology (No. 2021ZD0303200). Xiang-Dong Chen was supported by Key Research and Development Program of Anhui Province (No. 2022b13020006).
\bmsection{Disclosures}
The authors declare no conflicts of interest.
\bmsection{Data availability}
The pretraining data BioSR~\cite{qiao2021evaluation} and BioTISR~\cite{qiao2026neural} and the experimental test data from FlexSIM~\cite{soubies2024surpassing} and PCA-SIM~\cite{qian2023structured} are publicly available. The synthetic data, reconstruction results, and code are not publicly available at this time but may be obtained from the corresponding authors upon reasonable request.
\end{backmatter}

\balance
\bibliography{ref}

\bibliographyfullrefs{ref}

\end{document}